\documentclass[aip,pop]{revtex4-1}

\usepackage{graphicx}
\usepackage{natbib}

\usepackage{graphics}
\usepackage{epsfig}
\usepackage{bm}
\usepackage{amsmath,amssymb}
\usepackage{stmaryrd}
\usepackage{wasysym}
\usepackage[dvipsnames]{xcolor}

\def\wt#1{\widetilde{#1}}
\def\vb#1{\mbox{\boldmath$#1$}}
\def\pd#1#2{\frac{\partial #1}{\partial #2}}
\def\fd#1#2{\frac{\delta #1}{\delta #2}}
\def\wh#1{\widehat{#1}}
\def\bdot{\,\vb{\cdot}\,}
\def\btimes{\,\vb{\times}\,}

\def\bhat{\wh{{\sf b}}}
\def\cal#1{\mathcal{#1}}

\def\bhat{\wh{{\sf b}}}

\newcommand{\bc}{\begin{center}}
\newcommand{\ec}{\end{center}}
\newcommand{\bt}{\begin{tabbing}}
\newcommand{\et}{\end{tabbing}}
\newcommand{\be}{\begin{equation}}
\newcommand{\ee}{\end{equation}}
\newcommand{\ba}{\begin{eqnarray}}
\newcommand{\ea}{\end{eqnarray}}

\begin{document}

\title{Metriplectic bracket for guiding-center Vlasov-Maxwell-Landau theory}

\author{A.~J.~Brizard}
\affiliation{Department of Physics, Saint Michael's College, Colchester, VT 05439, USA}
\email{abrizard@smcvt.edu}
\author{H.~Sugama}
\affiliation{National Institute for Fusion Science, Toki 509-5292, Japan}
\affiliation{Department of Advanced Energy, University of Tokyo, Kashiwa 277-8561, Japan}

\begin{abstract}
The metriplectic formulation of collisional guiding-center Vlasov-Maxwell-Landau theory is presented. The guiding-center Landau collision operator, which describes collisions involving test-particle and field-particle guiding-center orbits, is represented in terms of a symmetric dissipative bracket involving functional derivatives of the guiding-center Vlasov phase-space density $F_{\rm gc}$ and the electromagnetic fields $({\bf D}_{\rm gc},{\bf B})$, where the guiding-center displacement vector ${\bf D}_{\rm gc} \equiv {\bf E} + 4\pi\,\vb{\sf P}_{\rm gc}$ is expressed in terms of the electric field ${\bf E}$ and the guiding-center polarization $\vb{\sf P}_{\rm gc}$. This dissipative Landau bracket conserves guiding-center energy-momentum and angular momentum, as well as satisfying a guiding-center H-theorem. 
\end{abstract}

\date{\today}

\maketitle

\section{Introduction}

Fluid and kinetic models are important tools used to study the complex dynamics of magnetized laboratory and astrophysical plasmas. A crucial element of these models is the clear separation of dissipationless and dissipative dynamics, which are represented by explicit evolution operators with important energy-momentum conservation laws.

A generic field-theoretical model found in plasma physics \cite{Morrison_2017} describes the time evolution of a set of field variables $\psi^{i}$ ($i = 1, 2, ..., N)$:
\begin{equation}
\pd{\psi^{i}}{t} \;=\; [\psi^{i},\;{\cal H}] \;+\; (\psi^{i},\;{\cal S}),
\label{eq:psi_dot}
\end{equation}
which is divided into a {\it dissipationless} (reversible) time evolution and {\it dissipative} (irreversible) time evolution, respectively. The dissipationless time evolution, on one hand, is generated by a Hamiltonian functional ${\cal H}[\vb{\psi}]$ and an antisymmetric bracket  \cite{Morrison_1982,Marsden_Weinstein_1982}
\begin{equation}
\left[{\cal F},\frac{}{}{\cal G}\right] \;=\; \int \fd{\cal F}{\psi^{i}}\;\wh{J}^{ij}[\vb{\psi}]\;\fd{\cal G}{\psi^{j}} \;=\; -\;\left[{\cal G},\frac{}{}{\cal F}\right],
\label{eq:bracket_Ham}
\end{equation}
which must satisfy the Jacobi property 
\begin{equation}
\left[ {\cal F},\frac{}{} [{\cal G},\;{\cal K}]\right] \;+\; \left[ {\cal G},\frac{}{} [{\cal K},\;{\cal F}]\right] \;+\; \left[ {\cal K},\frac{}{} [{\cal F},\;{\cal G}]\right] \;=\; 0
\label{eq:Jacobi_MV}
\end{equation}
that must be valid for any three functionals $({\cal F},{\cal G},{\cal K})$. The  {\it dissipative} time evolution, on the other hand, is generated by an entropy functional ${\cal S}[\vb{\psi}]$ and a symmetric dissipative bracket \cite{Kaufman_1984,Morrison:1984ca,Morrison:1986vw}
\begin{equation}
\left({\cal F},\frac{}{}{\cal G}\right) \;=\; \int \fd{\cal F}{\psi^{i}}\;\wh{Q}^{ij}[\vb{\psi}]\;\fd{\cal G}{\psi^{j}} \;=\; \left({\cal G},\frac{}{}{\cal F}\right).
\label{eq:bracket_diss}
\end{equation}
Here, the brackets \eqref{eq:bracket_Ham}-\eqref{eq:bracket_diss} are defined in terms of arbitrary functionals ${\cal F}[\vb{\psi}]$ and ${\cal G}[\vb{\psi}]$, and the integration domain is either physical space (for fluid models) and/or phase space (for kinetic models). In addition, summation over repeated indices is assumed, and the matrix elements $\wh{J}^{ij}[\vb{\psi}]$ and $\wh{Q}^{ij}[\vb{\psi}]$ typically involve partial derivatives of the functional derivatives $\delta(...)/\partial\psi^{i}$, which are defined through the Fr\'{e}chet derivative
\[ \left.\frac{d}{d\epsilon}{\cal F}[\vb{\psi} + \epsilon\,\delta\vb{\psi}]\right|_{\epsilon = 0} \;\equiv\; \int\fd{\cal F}{\psi^{i}}\;\delta\psi^{i}. \]
We note that the entropy functional ${\cal S}$ is a Casimir functional of the Hamiltonian bracket (i.e., $[{\cal S},\;{\cal G}] = 0$ for any arbitrary functional ${\cal G}$), while the Hamiltonian functional ${\cal H}$ is a dissipative invariant (i.e., $({\cal H},\;{\cal G}) = 0$ for any arbitrary functional ${\cal G}$). The entropy functional ${\cal S}$ and the dissipative bracket $(\;,\;)$ must also satisfy an H-Theorem\cite{Kaufman_Cohen_Brizard_2025}: $({\cal S},\,{\cal S}) \geq 0$, where $({\cal S},\,{\cal S}) = 0$ defines an equilibrium plasma state. The combination of the dissipationless (Hamiltonian) structure $\{{\cal H}; [\,,\,]\}$ and the dissipative structure $\{{\cal S}; (\,,\,)\}$ forms what is referred to as the {\it metriplectic} structure for the field equations\cite{Kaufman_1984,Morrison:1984ca,Morrison:1986vw}. This metriplectic structure can be implemented in advanced numerical algorithms \cite{Kraus_Hirvijoki_2017} that simultaneously evolve particle distributions and electromagnetic fields. 

The Hamiltonian structure for the original Vlasov-Maxwell equations, which is expressed in terms of the Vlasov distribution $f({\bf x},{\bf p};t)$ and the Maxwell electromagnetic fields ${\bf E}({\bf x},t)$ and ${\bf B}({\bf x},t)$, is partially expressed in terms of a noncanonical Poisson bracket $\{\;,\;\}_{\bf z}$ on particle phase space ${\bf z} \equiv ({\bf x},{\bf p})$ that satisfies its own Jacobi property. The Hamiltonian structures of the gauge-free gyrokinetic Vlasov-Maxwell equations \cite{Burby_Brizard_Morrison_Qin:2015PhLA,Burby_Brizard_2019,Brizard_2021_gyVM} and the guiding-center Vlasov-Maxwell equations \cite{Brizard_2021_gcVM,Brizard_gcVM_2024} have also been derived recently, with their respective Hamiltonian brackets satisfying the Jacobi property \eqref{eq:Jacobi_MV}. The major difference between the gyrokinetic and the guiding-center formalisms is how the electromagnetic fields are represented. In the gyrokinetic formalism, on the one hand, the electromagnetic fields are separated into time-independent background fields $({\bf E}_{0},{\bf B}_{0})$ and time-dependent fluctuating perturbed fields $({\bf E}_{1},{\bf B}_{1})$, and the Hamiltonian structure relies entirely on the perturbed fields $({\bf E}_{1},{\bf B}_{1})$. Within this gyrokinetic formalism, the conservation law of toroidal angular momentum depends on the symmetry properties of the background fields $({\bf E}_{0},{\bf B}_{0})$ \cite{Brizard_2021_exact}. In the guiding-center formalism, on the other hand, the electromagnetic fields are not separated into background and fluctuating components and, consequently, the conservation of energy-momentum and toroidal angular momentum are exact conservation laws, where the angular momentum conservation law follows from the symmetry of the guiding-center Vlasov-Maxwell stress tensor \cite{Brizard_Tronci_2016}. This explicit symmetry arises from important contributions of the guiding-center magnetization, which introduces a more complex Hamiltonian structure that is absent in the Hamiltonian gyrokinetic formalism.

More recently, the metriplectic structure of the gauge-free gyrokinetic Vlasov-Maxwell-Landau equations \cite{Hirvijoki_Burby_2020,Hirvijoki_2022} was derived based on the Hamiltonian structure of the gauge-free gyrokinetic Vlasov-Maxwell equations \cite{Burby_Brizard_2019,Brizard_2021_gyVM}. Once again, the toroidal angular momentum conservation properties of the gyrokinetic Landau collision operator rely on the axisymmetry of the background magnetic field ${\bf B}_{0}$. In the present work, we explore the metriplectic structure for the guiding-center Vlasov-Maxwell-Landau equations, in which the collisional effects are introduced in the guiding-center Vlasov equation through the guiding-center Landau collision operator. Here, we will see how guiding-center polarization and magnetization play a crucial role in the metriplectic structure of the guiding-center Vlasov-Maxwell-Landau equations. Lastly, the guiding-center metriplectic structure will preserve the energy-momentum and the toroidal angular momentum conservation laws, without any symmetry assumptions.

\section{Metriplectic bracket for Vlasov-Maxwell-Landau Theory}

The kinetic paradigm for the time evolution of a magnetized plasma is represented by the Vlasov-Maxwell-Landau theory, in which the Vlasov-Maxwell equations possess a Hamiltonian-bracket structure and the Landau collision operator can be represented in terms of a dissipative bracket and the Gibbs entropy functional.

\subsection{Vlasov-Maxwell-Landau theory}

The Vlasov-Maxwell-Landau equations are expressed in terms of the collisional Vlasov-Landau equation
\begin{equation}
\pd{f}{t} \;=\; -\;\frac{d{\bf x}}{dt}\bdot\nabla f - \frac{d{\bf p}}{dt}\bdot\pd{f}{\bf p} + \sum^{\prime}\;{\cal C}[f;f^{\prime}],
\label{eq:f_dot}
\end{equation}
which describes the time evolution of the distribution $f({\bf z},t)$ of test charged particles (with mass $m$ and charge $q$), defined in terms of the noncanonical phase-space coordinates ${\bf z} = ({\bf x},{\bf p})$, with Jacobian ${\cal J} = 1$. The two-particle Landau collision operator ${\cal C}[f;f^{\prime}]$ in Eq.~\eqref{eq:f_dot} is expressed in Landau and Fokker-Planck forms, respectively, as \cite{Helander_Sigmar} 
\begin{eqnarray}
 {\cal C}[f;f^{\prime}] &=& -\;\pd{}{\bf p} \bdot\left[ \int_{{\bf z}^{\prime}} \mathbb{Q}({\bf z},{\bf z}^{\prime})\bdot\left( f\;\pd{f^{\prime}}{{\bf p}^{\prime}} \;-\; f^{\prime}\;\pd{f}{\bf p}\right) \right] \nonumber \\
  &\equiv& -\;\pd{}{\bf p}\bdot\left( f\;{\bf K}[f'] \;-\; \mathbb{D}[f']\bdot\pd{f}{\bf p} \right),
 \label{eq:C_Landau}
 \end{eqnarray}
 where $f^{\prime}({\bf z}^{\prime},t)$ denotes the distribution of field particles (with mass $m^{\prime}$ and charge $q^{\prime}$), and summation over field-particle species is denoted as $\sum^{\prime}$. Here, the symmetric two-particle Landau tensor
 \begin{equation}
 \mathbb{Q}({\bf z},{\bf z}^{\prime}) \;\equiv\; Q_{0}\;\delta^{3}({\bf x}^{\prime} - {\bf x})\;\left( \frac{\mathbb{I}}{|{\bf w}|} \;-\; \frac{{\bf w}\,{\bf w}}{|{\bf w}|^{3}} \right)
 \label{eq:Q_def}
 \end{equation}
 is defined in terms of the relative velocity 
 \begin{equation}
 {\bf w} \;\equiv\;  \frac{d{\bf x}}{dt} \;-\; \frac{d{\bf x}^{\prime}}{dt} \;=\; {\bf v} \;-\; {\bf v}^{\prime}, 
 \label{eq:u_def}
 \end{equation}
the identity matrix $\mathbb{I}$, the constant $Q_{0} = 2\pi\,(qq^{\prime})^{2}\,\ln\Lambda$, and the presence of the delta function $\delta^{3}({\bf x}^{\prime} - {\bf x})$ guarantees that each collision occurs locally. We note that ${\bf w}$ is a null-eigenvector of the Landau matrix \eqref{eq:Q_def} (i.e., $\mathbb{Q}\bdot{\bf w} = 0$), which ensures exact energy conservation. 

The collisional Vlasov-Landau equations \eqref{eq:f_dot}-\eqref{eq:C_Landau} are complemented by the Maxwell equations
\begin{eqnarray}
\pd{\bf E}{t} &=& c\,\nabla\btimes{\bf B} \;-\; 4\pi \;\sum\,q\,\int_{\bf p} \frac{d{\bf x}}{dt}\;f, \label{eq:E_dot} \\
\pd{\bf B}{t} &=& -\;c\,\nabla\btimes{\bf E}, \label{eq:B_dot}
\end{eqnarray}
which describe the time evolution of the electric and magnetic fields ${\bf E}$ and ${\bf B}$, with the additional equations 
\begin{eqnarray}
\nabla\bdot{\bf E} &=& 4\pi \sum q\int_{\bf p} f, \label{eq:div_E} \\
\nabla\bdot{\bf B} &=& 0, \label{eq:div_B}
\end{eqnarray}
which can be used as initial conditions for Eqs.~\eqref{eq:E_dot}-\eqref{eq:B_dot}. In addition, the equations of motion in Eqs.~\eqref{eq:f_dot}-\eqref{eq:E_dot} are expressed in Hamiltonian form as
\begin{equation}
\frac{dz^{\alpha}}{dt} \;=\; \left\{ z^{\alpha},\; K\right\} \;+\; q\,{\bf E}\bdot\left\{ {\bf x},\; z^{\alpha}\right\},
\label{eq:z_dot}
\end{equation}
where $K \equiv |{\bf p}|^{2}/2m$ denotes the kinetic energy and the single-particle Poisson bracket is
\begin{equation}
\left\{ f,\frac{}{} g\right\} = \nabla f\bdot\pd{g}{\bf p} - \pd{f}{\bf p}\bdot\nabla g + \frac{q}{c}{\bf B}\bdot\pd{f}{\bf p}\btimes\pd{g}{\bf p}.
\label{eq:PB_def}
\end{equation}
The single-particle Poisson bracket \eqref{eq:PB_def} is antisymmetric, $\{f,\; g\} = -\,\{g,\;f\}$, and it satisfies the Jacobi property
\begin{equation}
\left\{ f,\frac{}{}\{g, h\}\right\} \;+\; \left\{ g,\frac{}{}\{h, f\}\right\} \;+\; \left\{ h,\frac{}{}\{f, g\}\right\} \;=\; 0,
\label{eq:PB_Jacobi}
\end{equation}
which is subject to the condition $\nabla\bdot{\bf B} = 0$, and is valid for any phase-space functions $(f,g,h)$. We note that the separation between the kinetic energy and the electric field in Eq.~\eqref{eq:z_dot} plays a crucial role in ensuring that the resulting Vlasov-Maxwell functional bracket \eqref{eq:VM_bracket} is antisymmetric.

Lastly, we note that, using Eq.~\eqref{eq:div_E}, the divergence of Eq.~\eqref{eq:E_dot} yields the charge conservation law
\begin{equation}
\pd{\varrho}{t} \;\equiv\; \sum q\,\int_{\bf p} \pd{f}{t} \;=\; -\,\nabla\bdot\left(\sum q\int_{\bf p} \frac{d{\bf x}}{dt}\;f \right) \;\equiv\; -\;\nabla\bdot{\bf J},
\label{eq:q_cons}
\end{equation}
where we used the fact that the collision operator \eqref{eq:C_Landau} is an exact momentum-divergence operator: $\int_{\bf p}{\cal C}[f;f^{\prime}] = 0$.

\subsection{Metriplectic structure of Vlasov-Maxwell-Landau theory}

We will now unify the Vlasov-Landau collisional kinetic equation \eqref{eq:f_dot}-\eqref{eq:C_Landau}  and  the Maxwell equations \eqref{eq:E_dot}-\eqref{eq:B_dot}  within a metriplectic bracket structure that combines dissipationless Hamiltonian evolution with Landau collisional evolution \cite{Kaufman_1984,Morrison:1984ca,Morrison:1984wu,Morrison:1986vw,Kraus_Hirvijoki_2017}. The evolution of an arbitrary Vlasov-Maxwell-Landau functional ${\cal F}[f,{\bf E},{\bf B}]$ is expressed in metriplectic form as
\begin{eqnarray}
\pd{\cal F}{t} &=& \sum \int_{\bf z}\pd{f}{t}\,\fd{\cal F}{f} \;+\; \int_{\bf x} \left(\pd{\bf E}{t}\bdot\fd{\cal F}{\bf E} + \pd{\bf B}{t}\bdot\fd{\cal F}{\bf B}\right) \nonumber \\
&\equiv& \left[{\cal F},\frac{}{} {\cal H}\right] \;+\; \left({\cal F},\frac{}{} {\cal S}\right),
\label{eq:F_dot_def}
\end{eqnarray}
where the Hamiltonian functional ${\cal H}$, and its associated antisymmetric Poisson bracket $[\;,\;]$, describe the collisionless evolution of ${\cal F}$, while the entropy functional ${\cal S}$, and its associated symmetric dissipative bracket $(\;,\;)$, describe the collisional evolution of ${\cal F}$. 

First, the Vlasov-Maxwell Poisson bracket between two arbitrary functionals $\cal F$ and $\cal G$ is expressed as \cite{Morrison_1980,Morrison_1982,Marsden_Weinstein_1982}
\begin{eqnarray}
\left[{\cal F},\frac{}{}{\cal G}\right] & \equiv & \sum \int_{\bf z}\;f \left\{ \fd{{\cal F}}{f},\; \fd{{\cal G}}{f} \right\} \label{eq:MV_PB} \;+\; 4\pi \int_{\bf x} \fd{{\cal F}}{{\bf E}} \bdot \left( c\,\nabla\btimes\fd{{\cal G}}{{\bf B}} - \sum\int_{\bf p} q\;f \left\{ {\bf x},\;\fd{{\cal G}}{f}\right\} \right)  \nonumber \\
 & - & 4\pi \int_{\bf x} \fd{{\cal G}}{{\bf E}} \bdot \left( c\,\nabla\btimes\fd{{\cal F}}{{\bf B}} - \sum\int_{\bf p} q\;f \left\{ {\bf x},\;\fd{{\cal F}}{f}\right\} \right),
 \label{eq:VM_bracket}
\end{eqnarray}
and the Hamiltonian functional is
\begin{equation}
{\cal H}[f,{\bf E},{\bf B}] \;\equiv\; \sum\int_{\bf z}K\;f \;+\; \int_{\bf x} \frac{1}{8\pi} \left( |{\bf E}|^{2} \;+\; |{\bf B}|^{2} \right).
\label{eq:H_def}
\end{equation}
The functional bracket \eqref{eq:MV_PB} satisfies the standard Poisson-bracket properties: (i) the antisymmetry property $[{\cal F},{\cal G}] = -\,[{\cal G},{\cal F}]$, (ii) the Leibniz property $[{\cal F}\,{\cal G},\;{\cal K}] = [{\cal F},{\cal K}]\,{\cal G} + {\cal F}\,[{\cal G}, {\cal K}]$, and (iii) the Jacobi property \eqref{eq:Jacobi_MV}, which is inherited from the Jacobi property \eqref{eq:PB_Jacobi} of the single-particle Poisson bracket \eqref{eq:PB_def}.

Next, the dissipative bracket in Eq.~\eqref{eq:F_dot_def} is expressed in the Landau form as \cite{Kaufman_1984,Kraus_Hirvijoki_2017}
\begin{equation}
\left({\cal F},\frac{}{}{\cal G}\right) \;\equiv\; \frac{1}{2}\int_{z, z'} \vb{\Gamma}[{\cal F}]\bdot\mathbb{Q}\bdot\vb{\Gamma}[{\cal G}],
\label{eq:DB_def}
\end{equation}
where the notation
\[ \int_{z, z'} (\cdots)  \equiv \sum\int_{\bf z} f({\bf z})\;\sum^{\prime}\int_{\bf z'}f^{\prime}({\bf z'})(\cdots) \]
involves a double particle phase-space integration over the test-particle coordinate ${\bf z}$ and the field-particle coordinate 
${\bf z}^{\prime}$, and the antisymmetric vector-valued function
\begin{eqnarray}
\vb{\Gamma}[{\cal F}]({\bf z};{\bf z'}) &\equiv& \pd{}{{\bf p}}\left(\fd{\cal F}{f}\right) - \pd{}{{\bf p'}}\left(\fd{\cal F}{f'}\right) \;=\; \left\{ {\bf x}, \fd{\cal F}{f}\right\} - \left\{ {\bf x'}, \fd{\cal F}{f'}\right\}^{\prime} = -\;\vb{\Gamma}[{\cal F}]({\bf z'};{\bf z})
\label{eq:gamma_particle}
\end{eqnarray}
involves functional derivatives with respect to the test-particle and field-particle distributions $f$ and $f'$. 

We now note that the Gibbs entropy functional (the Boltzmann constant is omitted for simplicity)
\begin{equation}
{\cal S} \;\equiv\; -\;\sum\int_{\bf z} f({\bf z})\;\ln f({\bf z})
\label{eq:S_def}
\end{equation}
is a Casimir functional of the Hamiltonian bracket \eqref{eq:MV_PB}, i.e., the bracket expression
\begin{eqnarray}
[{\cal S}, {\cal G}] &=& \sum\int_{\bf z} f\left\{ \fd{\cal S}{f}, \fd{\cal G}{f}\right\} + 4\pi \sum q \int_{\bf z} \fd{\cal G}{\bf E}\bdot f\,\left\{{\bf x}, \fd{\cal S}{f}\right\} \nonumber \\
 &=& -\,\sum\int_{\bf z} \left\{ f, \fd{\cal G}{f}\right\} - 4\pi \sum q \int_{\bf z} \fd{\cal G}{\bf E}\bdot\pd{f}{\bf p} \;=\; 0
\end{eqnarray}
vanishes for all functionals ${\cal G}$, since each integral vanishes separately since they involve exact divergences. Next, we calculate
\begin{eqnarray}
\vb{\Gamma}[{\cal S}]({\bf z};{\bf z'}) &=& \pd{}{{\bf p}}\left(-\ln f - 1\right) \;-\; \pd{}{{\bf p'}}\left(-\ln f' - 1\right) \;=\; -\,\pd{\ln f}{{\bf p}} \;+\; \pd{\ln f'}{{\bf p'}},
\end{eqnarray}
so that the dissipative Landau bracket \eqref{eq:DB_def} yields the collision operator ${\cal C}[f;f^{\prime}]$:
 \begin{eqnarray}
 \left({\cal F},\frac{}{}{\cal S}\right) &=& \frac{1}{2}\int_{z, z'} \vb{\Gamma}[{\cal F}]\bdot\mathbb{Q}\bdot\vb{\Gamma}[{\cal S}] \nonumber \\
 &=& \sum\int_{\bf z} \pd{}{\bf p}\left(\fd{\cal F}{f}\right)\bdot\sum^{\prime}\int_{{\bf z}^{\prime}}\mathbb{Q}\bdot\left( -\,f^{\prime}\;\pd{f}{\bf p} \;+\; f\;\pd{f^{\prime}}{{\bf p}^{\prime}}\right) \nonumber \\
   &\equiv& \sum\int_{\bf z} \pd{}{\bf p}\left(\fd{\cal F}{f}\right)\bdot \sum^{\prime}\left(-\;\mathbb{D}[f']\bdot\pd{f}{\bf p} \;+\; {\bf K}[f']\;f \right) \nonumber \\
 &\equiv& \sum\int_{\bf z} \pd{}{\bf p}\left(\fd{\cal F}{f}\right)\bdot \sum^{\prime}f\;{\bf S}[f'] \;=\; \sum\int_{\bf z} \fd{\cal F}{f}\;\sum^{\prime}{\cal C}[f;f^{\prime}].
\label{eq:FP-def} 
\end{eqnarray}
Here, the collision operator is given in Fokker-Planck form as
\begin{eqnarray}
{\cal C}[f;f'] &=& -\,\pd{}{\bf p}\bdot\left(f\frac{}{}{\bf S}[f']\right) \;\equiv\; -\,\pd{}{\bf p}\bdot\left({\bf K}[f']\;f \;-\; \mathbb{D}[f']\bdot\pd{f}{\bf p}\right),
 \label{eq:C_FP}
\end{eqnarray}
where the Fokker-Planck friction vector and the diffusion tensor are defined, respectively, as
\begin{eqnarray}
{\bf K}[f'] &=& \int_{{\bf z}^{\prime}}\mathbb{Q}({\bf z},{\bf z}^{\prime}) \bdot\pd{f^{\prime}}{{\bf p}^{\prime}}, \\
\mathbb{D}[f'] &=& \int_{{\bf z}^{\prime}}\mathbb{Q}({\bf z},{\bf z}^{\prime}) \;f^{\prime}, 
\end{eqnarray}
are Landau functionals of the field-particle distribution $f'$. Here, by introducing an arbitrary field point ${\bf x}_{0}$ in physical space, we note the symmetry property
\begin{equation}
\int_{\bf z}f({\bf z})\,\delta^{3}({\bf x} - {\bf x}_{0})\;{\bf S}[f']({\bf z}) \;\equiv\; -\;\int_{{\bf z}'} f'({\bf z}')\,\delta^{3}({\bf x}' - {\bf x}_{0})\;{\bf S}[f]({\bf z}'),
\label{eq:FPL_symmetry}
\end{equation}
which follows from the identity
\begin{eqnarray}
\int_{\bf z} f({\bf z})\;\delta^{3}({\bf x} - {\bf x}_{0})\,{\bf K}[f']({\bf z}) &=& \int_{\bf z} f({\bf z})\;\delta^{3}({\bf x} - {\bf x}_{0})\left(\int_{{\bf z}'}\mathbb{Q}({\bf z},{\bf z}^{\prime})\bdot\pd{f'({\bf z}')}{{\bf p}'}\right) \nonumber \\
 &=& \int_{{\bf z}'} \delta^{3}({\bf x}' - {\bf x}_{0})\left(\int_{\bf z} f({\bf z})\frac{}{}\mathbb{Q}({\bf z},{\bf z}^{\prime})\right)\bdot\pd{f'({\bf z}')}{{\bf p}'} \nonumber \\
  &\equiv& \int_{{\bf z}'} \delta^{3}({\bf x}' - {\bf x}_{0})\;\mathbb{D}[f]({\bf z}')\;\bdot\pd{f'({\bf z}')}{{\bf p}'},
\end{eqnarray}
where we used the symmetry and locality properties of the two-particle Landau tensor \eqref{eq:Q_def}. Because of the symmetry property \eqref{eq:FPL_symmetry}, we conclude that the local collisional force 
\begin{equation}
{\bf F}^{\sf C}({\bf x}_{0}) \;\equiv\; \int_{\bf z}{\bf p}\,\delta^{3}({\bf x} - {\bf x}_{0})\,\sum^{\prime}{\cal C}[f;f']({\bf z}) \;=\; \int_{\bf z}f({\bf z})\,\delta^{3}({\bf x} - {\bf x}_{0})\;\sum^{\prime}{\bf S}[f']({\bf z})
\label{eq:Force_coll}
\end{equation}
satisfies the momentum-conservation property $\sum{\bf F}^{\sf C}({\bf x}_{0}) = 0$, which follows from the locality of the two-particle Landau tensor \eqref{eq:Q_def}.

The conservation properties of the dissipative Landau bracket \eqref{eq:DB_def} easily follow from its form and the properties of the Landau kernel \eqref{eq:Q_def}. First, the particle-number functional 
${\cal N} = \int_{\bf z}f$ is trivially conserved since $\delta{\cal N}/\delta f = 1$ is a constant. Next, the conservation of the momentum functional
\begin{equation}
\vb{\cal P} \;=\; \sum\int_{\bf z}{\bf p}\,f \;+\; \int_{\bf x}\left({\bf E}\btimes{\bf B}\right)/4\pi c
\label{eq:P_VM}
\end{equation}
is also trivial since $\Gamma^{i}[{\cal P}^{j}] = \delta^{ij} - \delta^{ij} = 0$, while the conservation of the energy (Hamiltonian) functional 
\begin{equation}
{\cal H} \;=\;  \sum\int_{\bf z}\frac{|{\bf p}|^{2}}{2m}\,f \;+\; \int_{\bf x}\left(|{\bf E}|^{2} + |{\bf B}|^{2}\right)/8\pi 
\label{eq:H_VM}
\end{equation}
follows from the definition
\begin{eqnarray}
\vb{\Gamma}[{\cal H}]({\bf z};{\bf z}') &=& \left\{ {\bf x}, \fd{\cal H}{f}\right\} \;-\; \left\{ {\bf x'}, \fd{\cal H}{f'}\right\}^{\prime} \;=\; \frac{d{\bf x}}{dt} \;-\; \frac{d{\bf x}^{\prime}}{dt} \;\equiv\; {\bf v} \;- \; {\bf v}^{\prime} \;=\; {\bf w},
\label{eq:Gamma_H}
\end{eqnarray}
and the fact that ${\bf w}$ is a null-eigenvector of $\mathbb{Q}$: ${\bf w}\bdot\mathbb{Q} = 0$. Lastly, the dissipative Landau bracket \eqref{eq:DB_def} satisfies an H-theorem:
\begin{eqnarray}
\pd{\cal S}{t} &=& \left({\cal S},\frac{}{} {\cal S}\right) = \frac{1}{2}\int_{z,z^{\prime}}\;\vb{\Gamma}[{\cal S}]\bdot\mathbb{Q}\bdot\vb{\Gamma}[{\cal S}] = \frac{Q_{0}}{2}\int_{z, z^{\prime}}\frac{\delta^{3}({\bf x} - {\bf x}^{\prime})}{|{\bf w}|^{3}}\;\left|\vb{\Gamma}[{\cal S}]\btimes\frac{}{}{\bf w}\right|^{2} \;\geq\; 0,
 \label{eq:H_Theorem}
\end{eqnarray}
where thermal equilibrium $({\cal S},\;{\cal S}) = 0$ is defined by the condition $\vb{\Gamma}[{\cal S}]\btimes{\bf w} = 0$, i.e., when $f$ and $f^{\prime}$ are equal-temperature local Maxwellian distributions: $\vb{\Gamma}[{\cal S}] = {\bf w}/T$ (the temperature $T$ is given here in energy units). 

\section{\label{sec:Hgc_VM}Hamiltonian Guiding-center Vlasov-Maxwell Theory}

We now briefly review the Hamiltonian formulation of the guiding-center Vlasov-Maxwell equations, which was recently presented in our previous work\cite{Brizard_gcVM_2024} in terms of an antisymmetric Hamiltonian bracket $[\;,\;]_{\rm gc}$ and a Hamiltonian functional ${\cal H}_{\rm gc}$. In the next sections, we will consider the guiding-center Fokker-Planck collision operator (Sec.~\ref{sec:gc_FP}) derived by the guiding-center transformation of the Fokker-Planck collision operator and the dissipative-bracket formulation of a guiding-center Landau collision operator (Sec.~\ref{sec:gcL})  that describes collisions involving particle orbits represented in guiding-center phase space.

The variational formulation of guiding-center Lagrangian single-particle dynamics\cite{Cary_Brizard_2009} is based on a preliminary transformation $({\bf x}, {\bf p}) \rightarrow ({\bf x}, p_{\|0} = {\bf p}\bdot\bhat,\mu_{0} = |{\bf p}_{\bot}|^{2}/(2m B),\zeta_{0})$ to local particle phase-space coordinates, followed by a near-identity transformation to guiding-center phase-space coordinates $({\bf x}, p_{\|0},\mu_{0},\zeta_{0}) \rightarrow ({\bf X},p_{\|},\mu,\zeta)$, where ${\bf X}$ denotes the guiding-center position, $p_{\|}$ denotes the guiding-center parallel momentum, and $(\mu,\zeta)$ denotes the fast gyromotion canonical pair. This phase-space transformation induces Lie-transform guiding-center pull-back and push-forward operators ${\sf T}_{\rm gc}$ and ${\sf T}_{\rm gc}^{-1}$ that are involved in the transformation of the Vlasov particle distribution $f$ to the guiding-center Vlasov distribution $F \equiv {\sf T}_{\rm gc}^{-1}f$, while the guiding-center Poisson bracket is defined as $\{F,\;G\}_{\rm gc} \equiv {\sf T}_{\rm gc}^{-1}(\{ {\sf T}_{\rm gc}F,\; {\sf T}_{\rm gc}G\})$, which is guaranteed to satisfy the standard Poisson-bracket properties. In addition, the guiding-center time evolution operator $d_{\rm gc}/dt$ is defined in terms of the identity
\begin{equation}
{\sf T}_{\rm gc}^{-1}\left(\frac{df}{dt}\right) = \left({\sf T}_{\rm gc}^{-1}\frac{d}{dt}{\sf T}_{\rm gc}\right)F \equiv \frac{d_{\rm gc}F}{dt},
\label{eq:d_gc}
\end{equation}
which holds for an arbitrary function $f \equiv {\sf T}_{\rm gc}F$ and its guiding-center push-forward $F \equiv {\sf T}_{\rm gc}^{-1}f$. 

We note that the guiding-center operators ${\sf T}_{\rm gc}$ and ${\sf T}_{\rm gc}^{-1}$ are expressed in terms of expansions in powers of the mass-to-charge ratio $m/q$\cite{Kulsrud_1983}. In the present work, we will use the charge normalization $q \rightarrow q/\epsilon$ in order to explicitly display the guiding-center ordering parameter $\epsilon$ whenever it appears in an expression derived by guiding-center transformation. We note that, as an ordering parameter, it is not expected to be small {\it per se} and final guiding-center expressions may be truncated at any arbitrary order before setting it back to $\epsilon = 1$.

It is possible to construct a self-consistent system of guiding-center Vlasov-Maxwell equations, which describes the evolution of the guiding-center Vlasov distribution $F$ coupled with the Maxwell evolution of the electric and magnetic fields $({\bf D}_{\rm gc},
{\bf B})$. The variational (Lagrangian) formulations of the guiding-center Vlasov-Maxwell equations were presented by Pfirsch and Morrison\cite{Pfirsch_Morrison_1985} and more recently by Brizard and Tronci\cite{Brizard_Tronci_2016}, which also included a derivation of exact conservation laws for energy-momentum and angular momentum through the Noether method. 

\subsection{Guiding-center Lagrangian dynamics}

First, the guiding-center single-particle Lagrangian for a charged particle moving in a reduced phase space, with guiding-center position ${\bf X}$ and guiding-center parallel momentum $p_{\|}$, is expressed as
\begin{eqnarray}
L_{\rm gc} &=& \left(\frac{q}{\epsilon\,c}\,{\bf A} + p_{\|}\,\bhat\right)\bdot\frac{d_{\rm gc}{\bf X}}{dt} \;+\; \epsilon\,\mu\,\frac{B}{\Omega}\;\frac{d_{\rm gc}\zeta}{dt} \;-\; \left(\epsilon^{-1}q\,\Phi \;+\frac{}{} K_{\rm gc}\right),
\label{eq:Lag_gc}
\end{eqnarray}
where the guiding-center magnetic moment $\mu$ is a guiding-center invariant (i.e., because the guiding-center Lagrangian is independent of the gyroangle $\zeta$, $\partial L_{\rm gc}/\partial\zeta = 0$, the canonically-conjugate gyroaction $J = \mu\,B/\Omega \equiv \epsilon^{-1}\partial L_{\rm gc}/\partial\dot{\zeta}$ is an invariant) and the guiding-center kinetic energy is defined as\cite{Brizard_gcVM_2024}
\begin{equation}
K_{\rm gc}({\bf X},p_{\|};\mu) \;\equiv\; p_{\|}^{2}/2m \;+\; \mu\,B \;-\; \frac{m}{2}\,\left|{\bf E}\btimes c\bhat/B\right|^{2},
\label{eq:K_gc}
\end{equation}
where the last term represents the ponderomotive guiding-center Hamiltonian.  The guiding-center Lagrangian \eqref{eq:Lag_gc} is first calculated by considering the relative particle dynamics in a local frame moving with the $E\times B$ velocity, ${\bf E}\btimes c\bhat/B$, and then carrying a guiding-center transformation\cite{Brizard_gcVM_2024} that removes the $E\times B$ momentum ${\bf E}\btimes q\bhat/\Omega$ from the symplectic part $(\epsilon^{-1}q{\bf A}/c + p_{\|}\,\bhat)\bdot d_{\rm gc}{\bf X}/dt$ in Eq.~\eqref{eq:Lag_gc}. 

As a result of this guiding-center transformation, which asymptotically separates the gyromotion action-angle coordinates $(J,\zeta)$ from the reduced guiding-center phase-space coordinates $({\bf X},p_{\|})$, the guiding-center push-forward of the particle position ${\sf T}_{\rm gc}^{-1}{\bf x} = {\bf x}_{\rm gc}$ is expressed as (up to first order in $\epsilon$)
\begin{equation}
{\bf x}_{\rm gc}({\bf X},\mu,\zeta) \;=\; {\bf X} \;+\; \epsilon\,\vb{\rho}_{\rm gc}({\bf X},\mu,\zeta) ,
\label{eq:xgc_particle}
\end{equation}
where the guiding-center gyroradius
\begin{equation}
\vb{\rho}_{\rm gc} \;\equiv\; \vb{\rho}_{0}({\bf X},\mu,\zeta) \;+\; \frac{\bhat}{\Omega}\btimes\left({\bf E}\btimes\frac{c\bhat}{B}\right)
\label{eq:gyroradius}
\end{equation}
is expressed in terms of the lowest-order gyroangle-dependent gyroradius $\vb{\rho}_{0}({\bf X},\mu,\zeta)$ and the polarization displacement\cite{Cary_Brizard_2009,Brizard_gcVM_2024} $\vb{\rho}_{\rm E}({\bf X}) \equiv \bhat\btimes({\bf E}\btimes c\bhat/B)/\Omega$. 

The guiding-center equations of motion are derived from the guiding-center Lagrangian \eqref{eq:Lag_gc} as Euler-Lagrange equations in reduced guiding-center phase space $({\bf X},p_{\|})$:
\begin{eqnarray}
\frac{d_{\rm gc}p_{\|}}{dt}\;\bhat - \epsilon^{-1}q\,{\bf E}^{\star} - \frac{q}{\epsilon\,c}\frac{d_{\rm gc}{\bf X}}{dt}\btimes{\bf B}^{*} &=& -\;\nabla K_{\rm gc} \;=\; -\;\nabla \left(\mu\,B - \frac{mc^{2}}{2B^{2}}\,\left|{\bf E}\btimes\bhat\right|^{2}\right), \label{eq:EL_X} \\
\bhat\bdot\frac{d_{\rm gc}{\bf X}}{dt} &=& \pd{K_{\rm gc}}{p_{\|}} \;=\; \frac{p_{\|}}{m}, \label{eq:EL_p} 
\end{eqnarray}
where the effective electric and magnetic fields in Eq.~\eqref{eq:EL_X} are defined as
\begin{eqnarray}
{\bf E}^{\star} &\equiv & {\bf E} \;-\; \epsilon\,(p_{\|}/q)\,\partial\bhat/\partial t \;=\; {\bf E} \;+\; \epsilon\,\mathbb{P}_{\|}\bdot\nabla\btimes{\bf E}, \label{eq:E_star} \\
{\bf B}^{*} & \equiv & {\bf B} \;+\; \epsilon\,(p_{\|}c/q)\,\nabla\btimes\bhat. \label{eq:B_star}
\end{eqnarray}
Here, we assume the usual guiding-center ordering \cite{Kulsrud_1983} ${\bf E} = \epsilon\,E_{\|}\,\bhat + {\bf E}_{\bot}$ for the electric field, and we used the identity
\[ \pd{\bhat}{t} \;\equiv\; -\,\frac{c}{B}\,\left(\mathbb{I} \;-\; \bhat\,\bhat\right)\bdot\nabla\btimes{\bf E}, \]
which follows from Faraday's law \eqref{eq:B_dot}, where we introduced the symmetric dyadic tensor
\begin{equation}
 \mathbb{P}_{\|} \;\equiv\; \frac{p_{\|}c}{qB}\;\left(\mathbb{I} \;-\; \bhat\,\bhat\right),
 \label{eq:P_par}
 \end{equation}
which has units of a {\it parallel} gyroradius and is used to represent the guiding-center polarization contribution to guiding-center magnetization. 

The reduced guiding-center equations of motion can be obtained from the Euler-Lagrange equations \eqref{eq:EL_X}-\eqref{eq:EL_p}:
\begin{eqnarray}
\frac{d_{\rm gc}{\bf X}}{dt}  &=& \left\{{\bf X},\; K_{\rm gc}\right\}_{\rm rgc} \;+\; \epsilon^{-1}q\,{\bf E}^{\star}\bdot\{{\bf X},{\bf X}\}_{\rm rgc} \nonumber \\
 &=&\frac{{\bf B}^{*}}{B_{\|}^{*}}\;\pd{K_{\rm gc}}{p_{\|}} \;+\; \left(q\,{\bf E}^{\star} \;-\frac{}{} \epsilon\,\nabla K_{\rm gc}\right)\btimes\frac{c\bhat}{qB_{\|}^{*}}, \label{eq:X_dot_gc} \\
\frac{d_{\rm gc}p_{\|}}{dt} &=& \left\{p_{\|},\; K_{\rm gc}\right\}_{\rm rgc} \;+\; \epsilon^{-1}q\,{\bf E}^{\star}\bdot\{{\bf X}, p_{\|}\}_{\rm rgc} \nonumber \\
 &=& \left(\epsilon^{-1}q\,{\bf E}^{\star} \;-\frac{}{} \nabla K_{\rm gc}\right)\bdot\frac{{\bf B}^{*}}{B_{\|}^{*}}, \label{eq:p_dot_gc} 
 \end{eqnarray}
where $B_{\|}^{*} \equiv \bhat\bdot{\bf B}^{*}$ is the guiding-center Jacobian (ignoring a constant factor of mass $m$), which satisfies the guiding-center Liouville theorem
\begin{eqnarray}
\pd{B_{\|}^{*}}{t} &=& -\,c\,\bhat\bdot\nabla\times{\bf E}^{\star} \;-\; \epsilon^{-1}q\,{\bf B}^{*}\bdot\pd{{\bf E}^{\star}}{p_{\|}} \nonumber \\
 &=& -\,\nabla\bdot\left(B_{\|}^{*}\,\frac{d_{\rm gc}{\bf X}}{dt}\right) \;-\; \pd{}{p_{\|}}\left(B_{\|}^{*}\,\frac{d_{\rm gc}p_{\|}}{dt}\right),
\label{eq:Liouville_gc}
\end{eqnarray}
with the guiding-center equations \eqref{eq:X_dot_gc}-\eqref{eq:p_dot_gc}. We note that the noncanonical guiding-center Poisson bracket \cite{Cary_Brizard_2009}
\begin{equation}
\{ f,\; g\}_{\rm gc} \;\equiv\; \epsilon^{-1}\frac{\Omega}{B}\left(\pd{f}{\zeta}\pd{g}{\mu} - \pd{f}{\mu}\pd{g}{\zeta}\right) \;+\; \{f,\; g\}_{\rm rgc} 
\label{eq:PB_gc}
\end{equation}
is naturally separated into the gyromotion canonical bracket and the reduced guiding-center (rgc) noncanonical Poisson bracket 
\begin{equation}
\{f,\; g\}_{\rm rgc} \;\equiv\; \frac{{\bf B}^{*}}{B_{\|}^{*}}\bdot\left(\nabla f\;\pd{g}{p_{\|}} \;-\; \pd{f}{p_{\|}}\;\nabla g\right) \;-\; \frac{\epsilon\,c\bhat}{qB_{\|}^{*}}\bdot\nabla f\btimes\nabla g,
\label{eq:PB_rgc}
\end{equation} 
which operates on functions in reduced guiding-center phase space $({\bf X},p_{\|})$. We note that the guiding-center Poisson bracket \eqref{eq:PB_gc}
can also be expressed in divergence form as
\begin{equation}
\{ f,\; g\}_{\rm gc}  \;=\; \frac{1}{B_{\|}^{*}}\pd{}{Z^{\alpha}}\left(B_{\|}^{*}\frac{}{}f\;\left\{Z^{\alpha},\frac{}{} g\right\}_{\rm gc}\right),
\label{eq:PB_gc_div}
\end{equation}
and that it also satisfies the Jacobi property
\begin{equation}
\left\{ \{f, g\}_{\rm gc},\frac{}{} h\right\}_{\rm gc} + \left\{ \{g, h\}_{\rm gc},\frac{}{} f\right\}_{\rm gc} + \left\{ \{h, f\}_{\rm gc},\frac{}{} g\right\}_{\rm gc} = 0,
\label{eq:Jac_rgc}
\end{equation}
subject to the condition $\nabla\bdot{\bf B}^{*} = \nabla\bdot{\bf B} = 0$, which is satisfied by the definition \eqref{eq:B_star}.

\subsection{Guiding-center Vlasov-Maxwell equations}

Next, we introduce the guiding-center Vlasov-Maxwell equations \cite{Brizard_gcVM_2024}
\begin{eqnarray}
\pd{F_{\rm gc}}{t}  &=& -\,\nabla\bdot\left(F_{\rm gc}\,\frac{d_{\rm gc}{\bf X}}{dt}\right) - \pd{}{p_{\|}}\left(F_{\rm gc}\,\frac{d_{\rm gc}p_{\|}}{dt}\right), \label{eq:V_eq} \\
\pd{{\bf D}_{\rm gc}}{t} &=& c\,\nabla\btimes\left({\bf B} -\frac{}{} 4\pi\,\vb{\sf M}_{\rm gc}\right) - 4\pi \sum q\int_{P} F_{\rm gc}\,\frac{d_{\rm gc}{\bf X}}{dt} \nonumber \\
 &\equiv& c\,\nabla\times{\bf H}_{\rm gc} \;-\; 4\pi\,{\bf J}_{\rm gc}, \label{eq:Maxwell_eq} \\
\pd{\bf B}{t} &=& -\,c\,\nabla\btimes{\bf E}, \label{eq:Faraday_eq} 
\end{eqnarray}
where the guiding-center phase-space density $F_{\rm gc} \equiv F\,B_{\|}^{*}$ is defined in terms of the guiding-center Vlasov function $F$ and the guiding-center Jacobian $B_{\|}^{*}$, and the guiding-center momentum integral $\int_{P} \equiv \int (2\pi\,m)\;dp_{\|}\,d\mu$ excludes the guiding-center Jacobian $B_{\|}^{*}$. We note that the electromagnetic fields $({\bf D}_{\rm gc},{\bf B})$ also satisfy the divergence equations
\begin{eqnarray}
\nabla\bdot{\bf D}_{\rm gc} &=& 4\pi\,\sum q\;\int_{P} F_{\rm gc} \;\equiv\; 4\pi\,\varrho_{\rm gc}, \label{eq:div_D} \\
\nabla\bdot{\bf B} &=& 0, \label{eq:div_B}
\end{eqnarray}
where the guiding-center charge density $\varrho_{\rm gc}$ and the guiding-center current density ${\bf J}_{\rm gc}$ satisfy the charge conservation law
\begin{equation}
\pd{\varrho_{\rm gc}}{t} \;=\; \sum q\,\int_{P}\pd{F_{\rm gc}}{t} \;=\; -\,\nabla\bdot{\bf J}_{\rm gc} \;=\; -\,\nabla\bdot\left(\sum q \int_{P} F_{\rm gc}\;\frac{d_{\rm gc}{\bf X}}{dt}\right),
\label{eq:charge} 
\end{equation}
where we used the guiding-center Vlasov equation \eqref{eq:V_eq}.

In Eq.~\eqref{eq:Maxwell_eq}, the guiding-center electromagnetic fields ${\bf D}_{\rm gc} \equiv {\bf E} + 4\pi\,\vb{\sf P}_{\rm gc}$ and ${\bf H}_{\rm gc} \equiv {\bf B} - 4\pi\,\vb{\sf M}_{\rm gc}$ are defined in terms of the guiding-center polarization and magnetization \cite{Brizard_gcVM_2024}
\begin{eqnarray}
\vb{\sf P}_{\rm gc} &\equiv& \sum \int_{P} F_{\rm gc}\pd{L_{\rm gc}}{\bf E} \;=\; \sum\int_{P} F_{\rm gc}\;\vb{\pi}_{\rm E}, 
\label{eq:P_def} \\
\vb{\sf M}_{\rm gc} &\equiv& \sum\int_{P} F_{\rm gc}\pd{L_{\rm gc}}{\bf B} = \sum\int_{P} F_{\rm gc}\left( \vb{\mu}_{\rm gc} + \vb{\pi}_{\rm gc}\btimes\frac{p_{\|}\bhat}{mc}\right),
\label{eq:M_def}
\end{eqnarray}
where the guiding-center electric dipole moment $\vb{\pi}_{\rm E} \equiv q\,\vb{\rho}_{\rm E}$ is defined in terms of the polarization displacement \eqref{eq:gyroradius}, and the guiding-center dipole moments are defined as
\begin{eqnarray}
\vb{\pi}_{\rm gc} &\equiv& \frac{q\bhat}{\Omega}\btimes\frac{d_{\rm gc}{\bf X}}{dt}, \\
\vb{\mu}_{\rm gc} &\equiv& -\,\mu\,\bhat \;-\; \frac{\bf E}{B}\bdot\left(\bhat\,\vb{\pi}_{\rm E} \;+\frac{}{} \vb{\pi}_{\rm E}\,\bhat\right).
\end{eqnarray}
We note that the guiding-center magnetization \eqref{eq:M_def} combines the intrinsic magnetization due to $\vb{\mu}_{\rm gc}$ with the moving-electric-dipole magnetization 
\begin{equation}
\vb{\pi}_{\rm gc}\btimes\frac{p_{\|}\bhat}{mc} \;=\; \left( \frac{q\bhat}{\Omega}\btimes\frac{d_{\rm gc}{\bf X}}{dt}\right)\btimes\frac{p_{\|}\bhat}{mc} \;=\; \frac{q}{c}\,\mathbb{P}_{\|}\bdot\frac{d_{\rm gc}{\bf X}}{dt},
\label{eq:moving_gc}
\end{equation}
where we used the symmetric dyadic tensor \eqref{eq:P_par}. It was previously noted\cite{Sugama_2016,Brizard_Tronci_2016} that the guiding-center moving-electric-dipole magnetization plays a crucial role in constructing a symmetric guiding-center stress tensor, which is a requirement for the explicit proof of the guiding-center angular-momentum conservation law. Moreover, the definition of the guiding-center magnetization \eqref{eq:M_def}, which is derived from the entire guiding-center Lagrangian \eqref{eq:Lag_gc}, is different from the gyrokinetic magnetization, which is derived solely from the gyrocenter kinetic energy\cite{Burby_Brizard_Morrison_Qin:2015PhLA}.
 
\vspace*{0.1in}

\subsection{Guiding-center Hamiltonian bracket}

The guiding-center Hamiltonian functional is constructed from the energy conservation law for the guiding-center Vlasov-Maxwell equations \cite{Brizard_gcVM_2024}, and is expressed as
\begin{widetext}
\begin{eqnarray}
{\cal H}_{\rm gc}[F_{\rm gc},{\bf D}_{\rm gc},{\bf B}] &=& \sum\int_{\bf Z} F_{\rm gc}\;K_{\rm gc} \;+\;  \int_{\bf X} \left[ {\bf D}_{\rm gc}\bdot\frac{\bf E}{4\pi} \;-\; \frac{1}{8\pi}\left(|{\bf E}|^{2} - |{\bf B}|^{2}\right)\right] \nonumber \\
 &\equiv& \sum\int_{\bf Z} F_{\rm gc}\; \left( \mu\,B \;+\; \frac{p_{\|}^{2}}{2m}\right) + \int_{\bf X} \left[ \frac{|{\bf B}|^{2}}{8\pi} + \frac{1}{8\pi}\,{\bf D}_{\rm gc}\bdot\left(\overleftrightarrow{\vb{\varepsilon}_{\rm gc}}\right)^{-1}\bdot{\bf D}_{\rm gc}\right],
\label{eq:gc_H}
\end{eqnarray}
\end{widetext}
where $\int_{\bf Z} \equiv \int_{{\bf X},P}$ denotes integration over the 4+1 reduced guiding-center phase-space $({\bf X},p_{\|};\mu)$, while the guiding-center permittivity tensor 
\begin{equation}
\overleftrightarrow{\vb{\varepsilon}_{\rm gc}} \;\equiv\; \mathbb{I} \;+\; 4\pi\,\chi_{\rm gc}\;\left(\mathbb{I} \;-\; \bhat\,\bhat\right)
\label{eq:epsilon_gc}
\end{equation}
which is defined through the constitutive relation ${\bf D}_{\rm gc} \equiv \overleftrightarrow{\vb{\varepsilon}_{\rm gc}}\bdot{\bf E} = {\bf E} + 4\pi\,{\bf P}_{\rm gc}$, with ${\bf P}_{\rm gc} \equiv \chi_{\rm gc}\,{\bf E}_{\bot}$ defined through the scalar guiding-center susceptibility
\begin{equation}
\chi_{\rm gc}[F_{\rm gc},{\bf B}] \;=\; \sum\frac{mc^{2}}{B^{2}}\int_{P}\;F_{\rm gc}.
\label{eq:chi_gc}
\end{equation} 
We note that, for a typical magnetized fusion plasma, we find $4\pi\,\chi_{\rm gc} \equiv c^{2}/v_{\sf A}^{2} \gg 1$. The functional derivatives of the guiding-center Hamiltonian functional \eqref{eq:gc_H} are
\begin{equation}
\left( \begin{array}{c}
\delta{\cal H}_{\rm gc}/\delta F_{\rm gc} \\
\delta{\cal H}_{\rm gc}/\delta{\bf D}_{\rm gc} \\
\delta{\cal H}_{\rm gc}/\delta{\bf B}
\end{array} \right) \;=\; \left( \begin{array}{c}
K_{\rm gc} \\
{\bf E}/4\pi \\
{\bf B}/4\pi - \sum\int_{P}F_{\rm gc}\;\vb{\mu}_{\rm gc}
\end{array} \right),
\label{eq:delta_H}
\end{equation}
and we will need the modified functional derivative
 \begin{equation}
\frac{\delta^{\star}{\cal H}_{\rm gc}}{\delta{\bf D}_{\rm gc}} \equiv \fd{{\cal H}_{\rm gc}}{{\bf D}_{\rm gc}} + \epsilon\,\mathbb{P}_{\|}\bdot\nabla\btimes\fd{{\cal H}_{\rm gc}}{{\bf D}_{\rm gc}} = \frac{{\bf E}^{\star}}{4\pi},
\label{eq:delta_star}
\end{equation}
where ${\bf E}^{\star}$ is defined in Eq.~\eqref{eq:E_star}. We note that this modified functional derivative is a unique feature of the guiding-center Hamiltonian formalism\cite{Brizard_2021_gcVM}, which is connected to the guiding-center magnetization \eqref{eq:M_def}.

The guiding-center Vlasov-Maxwell bracket is expressed in terms of two arbitrary guiding-center functionals ${\cal F}$ and ${\cal G}$ as \cite{Brizard_gcVM_2024}
 \begin{eqnarray}
 \left[{\cal F},\frac{}{}{\cal G}\right]_{\rm gc} &=& \sum\int_{\bf Z} F_{\rm gc} \left\{ \fd{{\cal F}}{F_{\rm gc}} ,\; \fd{\cal G}{F_{\rm gc}} \right\}_{\rm gc}  \;+\; 4\pi c  \int_{\bf X} \left(\fd{\cal F}{{\bf D}_{\rm gc}}\bdot\nabla\btimes\fd{\cal G}{\bf B} - \fd{\cal G}{{\bf D}_{\rm gc}}\bdot\nabla\btimes\fd{\cal F}{\bf B} \right) \nonumber \\
  &&+\; \sum 4\pi q \int_{\bf Z} F_{\rm gc}\left( \frac{\delta^{\star}{\cal G}}{\delta{\bf D}_{\rm gc}}\bdot\left\{{\bf X},\; \fd{{\cal F}}{F_{\rm gc}} \right\}_{\rm gc} \;-\; \frac{\delta^{\star}{\cal F}}{\delta{\bf D}_{\rm gc}}\bdot\left\{{\bf X},\;\fd{\cal G}{F_{\rm gc}} \right\}_{\rm gc} \right) \nonumber \\
   &&+\;  \sum(4\pi q)^{2} \int_{\bf Z} F_{\rm gc} \left(\frac{\delta^{\star}{\cal F}}{\delta{\bf D}_{\rm gc}}\bdot\left\{{\bf X},\; {\bf X}\right\}_{\rm gc}\bdot \frac{\delta^{\star}{\cal G}}{\delta{\bf D}_{\rm gc}}\right).
  \label{eq:gcVM_bracket} 
 \end{eqnarray}
Here, we note that, since the class of guiding-center functionals is restricted by the condition that the functional derivative $\delta{\cal F}/\delta F_{\rm gc} \equiv \langle \delta{\cal F}/\delta F_{\rm gc}\rangle$ is gyroangle-independent, we may replace the guiding-center Poisson bracket \eqref{eq:PB_gc} with the reduced guiding-center Poisson bracket \eqref{eq:PB_rgc} without any consequences. In a recent paper \cite{Brizard_gcVM_proof}, it was shown that the guiding-center Vlasov-Maxwell bracket  \eqref{eq:gcVM_bracket} satisfies the Jacobi property \eqref{eq:Jacobi_MV}, which is inherited from the Jacobi property \eqref{eq:Jac_rgc} for the guiding-center Poisson bracket \eqref{eq:PB_gc}.

\subsection{Guiding-center Vlasov-Maxwell invariants}

We now discuss the conservation properties of the guiding-center Vlasov-Maxwell bracket \eqref{eq:gcVM_bracket}. Because of the antisymmetry property of the bracket, the guiding-center energy ${\cal H}_{\rm gc}$ is trivially conserved: 
$[{\cal H}_{\rm gc},{\cal H}_{\rm gc}]_{\rm gc} \equiv 0$. 

Next, the proofs of the conservation of the guiding-center Vlasov-Maxwell linear and angular momenta, respectively defined as\cite{Brizard_gcVM_2024}
\begin{eqnarray}
{\cal P}_{\rm gc}^{i} &=& \sum\int_{\bf Z} F_{\rm gc}\,p_{\|}\,b^{i} \;+\; \int_{\bf X} \wh{\imath}\bdot({\bf D}_{\rm gc}\btimes{\bf B})/4\pi c, \label{eq:P_gc_i} \\
{\cal P}_{{\rm gc}\varphi} &=& \sum\int_{\bf Z} F_{\rm gc}\,p_{\|}\,b_{\varphi} \;+\; \int_{\bf X} \pd{\bf X}{\varphi}\bdot({\bf D}_{\rm gc}\btimes{\bf B})/4\pi c, \label{eq:P_gc_phi}
\end{eqnarray}
where $b^{i} \equiv \wh{\imath}\bdot\bhat$ ($\wh{\imath}$ denotes an arbitrary constant translation direction) and $b_{\varphi} \equiv \bhat\bdot\partial{\bf X}/\partial\varphi$. Here, the divergenceless contravariant azimuthal derivative $\partial{\bf X}/\partial\varphi = \wh{\sf z}\btimes{\bf X}$, which is defined in terms of the unit vector $\wh{\sf z}$ directed along the axis of rotation, satisfies the vector identity ${\bf C}\bdot\nabla(\partial{\bf X}/\partial\varphi) = \wh{\sf z}\btimes{\bf C}$, for an arbitrary vector field ${\bf C}$. Once again, the conservation of the guiding-center angular momentum functional \eqref{eq:P_gc_phi} is based on the symmetry of the guiding-center Vlasov-Maxwell stress tensor\cite{Brizard_Tronci_2016}, which critically involves the guiding-center moving electric-dipole magnetization \eqref{eq:moving_gc}. 

An important property of the guiding-center Hamiltonian bracket \eqref{eq:gcVM_bracket} is that there exists a certain class of functionals known as Casimir invariant functionals ${\cal K}_{\rm gc}$ for which $[{\cal K}_{\rm gc}, 
{\cal G}]_{\rm gc} = 0$ for any functional ${\cal G}$. An important example, which will play a crucial role in our work, is the guiding-center Gibbs entropy functional \cite{Brizard_2021_gcVM}
\begin{equation}
{\cal S}_{\rm gc} \;=\; -\;\sum\int_{\bf Z} F_{\rm gc}\;\ln(F_{\rm gc}/B_{\|}^{*}).
\label{eq:S_gc}
\end{equation}
The functional derivatives of Eq.~\eqref{eq:S_gc} yield
\begin{eqnarray}
\fd{{\cal S}_{\rm gc}}{F_{\rm gc}} &=& -\;\ln F \;-\; 1, \\
\fd{{\cal S}_{\rm gc}}{\bf B} &=& \int_{P} \left( F\;\bhat - \frac{q}{c}\mathbb{P}_{\|}\bdot B_{\|}^{*}\{{\bf X},\; F\}_{\rm rgc} \right),
\end{eqnarray}
while $\delta{\cal S}_{\rm gc}/\delta{\bf D}_{\rm gc} = 0 = \delta^{\star}{\cal S}_{\rm gc}/\delta{\bf D}_{\rm gc}$, where $F \equiv F_{\rm gc}/B_{\|}^{*}$ denotes the guiding-center Vlasov function and we used the functional variation 
\begin{eqnarray} 
\delta B_{\|}^{*} &=& \delta{\bf B}^{*}\bdot\bhat \;+\; {\bf B}^{*}\bdot\delta\bhat  \\
 &=& \left[\delta{\bf B} \;+\frac{}{} \nabla\btimes(\epsilon\,\mathbb{P}_{\|}\bdot\delta{\bf B})\right]\bdot\bhat \;+\; \frac{q}{c}\,{\bf B}^{*}\bdot\pd{\mathbb{P}_{\|}}{p_{\|}}\bdot\delta{\bf B}.  \nonumber
\end{eqnarray}
When these functional derivatives are inserted in Eq.~\eqref{eq:gcVM_bracket}, we obtain
\begin{eqnarray}
[{\cal S}_{\rm gc},\; {\cal G}]_{\rm gc} &=& -\;\sum\int_{\bf Z} B_{\|}^{*} \left\{ F,\; \fd{\cal G}{F_{\rm gc}}\right\}_{\rm rgc} \;-\; 4\pi \sum q \int_{\bf Z} B_{\|}^{*} \frac{\delta^{\star}{\cal G}}{\delta{\bf D}_{\rm gc}}\bdot\{{\bf X},\; F\}_{\rm rgc} \nonumber \\
 &&-\; 4\pi c \sum\int_{\bf Z}\left( F\;\bhat - \frac{q}{c}\mathbb{P}_{\|}\vb{\cdot} B_{\|}^{*}\{{\bf X},\; F\}_{\rm rgc} \right)\vb{\cdot}\nabla\vb{\times}\fd{\cal G}{{\bf D}_{\rm gc}} \nonumber \\
   &=& -\;\sum\int_{\bf Z} B_{\|}^{*} \left\{ F,\; \fd{\cal G}{F_{\rm gc}}\right\}_{\rm rgc} \;+\; 4\pi c \sum\int_{\bf Z} \nabla\bdot\left(F\,\bhat\btimes\fd{\cal G}{{\bf D}_{\rm gc}}\right) \;=\; 0,
 \label{eq:SG_gc}
 \end{eqnarray}
 which vanishes for all functionals ${\cal G}$ since each remaining integral in Eq.~\eqref{eq:SG_gc} involves an exact divergence. 
 
A second example of a guiding-center Casimir functional, recently introduced by Sato and Morrison\cite{Sato_Morrison_2024,Sato_Morrison_2025}, is the total magnetic moment ${\cal M}_{\rm gc} \equiv \sum \int_{\bf Z} F_{\rm gc}\,\mu$, which yields the Hamiltonian bracket expression
 \begin{equation}
[{\cal M}_{\rm gc},\; {\cal G}]_{\rm gc} \;=\; \sum \int_{\bf Z} F_{\rm gc}\left\{ \mu,\; \fd{\cal G}{F_{\rm gc}}\right\}_{\rm gc} \;+\; \sum 4\pi\,q \int_{\bf Z} F_{\rm gc}\;\frac{\delta^{\star}{\cal G}}{\delta{\bf D}_{\rm gc}}\bdot\{{\bf X},\; \mu\}_{\rm gc}.
\label{eq:Casimir_mu}
\end{equation}
Since the magnetic moment $\mu$ satisfies the Casimir-like condition $\{\mu, g\}_{\rm gc} \equiv 0$ for any guiding-center function $g({\bf X},p_{\|};\mu)$ that is gyroangle-independent (i.e., $\partial g/\partial\zeta \equiv 0$), we then obtain the Casimir condition $[{\cal M}_{\rm gc},{\cal G}]_{\rm gc} \equiv 0$ for any guiding-center functional ${\cal G}$ for which $\delta{\cal G}/\delta F_{\rm gc} \equiv \langle\delta{\cal G}/\delta F_{\rm gc}\rangle$.  Sato and Morrison\cite{Sato_Morrison_2024,Sato_Morrison_2025} refer to the guiding-center functional ${\cal M}_{\rm gc}$ as an {\it interior} Casimir functional.
 
\section{\label{sec:gc_FP}Guiding-center Fokker-Planck Collision Operator}

When two-particle collisions are introduced into the guiding-center Vlasov equation \eqref{eq:V_eq}, the time evolution of the guiding-center phase-space density $F_{\rm gc}$ is now expressed in terms of the guiding-center Fokker-Planck equation
\begin{equation}
\frac{d_{\rm gc}F_{\rm gc}}{dt} \;\equiv\; \pd{F_{\rm gc}}{t} + \nabla\bdot\left(F_{\rm gc}\,\frac{d_{\rm gc}{\bf X}}{dt}\right) + \pd{}{p_{\|}}\left(F_{\rm gc}\,\frac{d_{\rm gc}p_{\|}}{dt}\right) \;=\; B_{\|}^{*}\,\sum^{\prime}{\cal C}_{\rm gc}[F; F'],
\end{equation}
where the guiding-center Fokker-Planck collision operator can be formally derived by Lie-transform perturbation methods as a collision operator in 4+1 dimensional space $({\bf X},p_{\|};\mu)$ \cite{Brizard_2004}
\begin{equation}
{\cal C}_{\rm gc}[F;F'] \equiv \left\langle {\sf T}_{\rm gc}^{-1}{\cal C}\left[{\sf T}_{\rm gc}\frac{}{}F; {\sf T}^{\prime}_{\rm gc}\frac{}{}F'\right] \right\rangle = -\,\frac{1}{B_{\|}^{*}}\pd{}{Z^{\alpha}}\left[B_{\|}^{*}\left( K_{\rm gc}^{\alpha}[F']\,F - 
D_{\rm gc}^{\alpha\beta}[F']\;\pd{F}{Z^{\beta}}\right)\right].
\label{eq:C_gc}
\end{equation}
Here, the test-particle and field-particle guiding-center distributions $F \equiv \langle {\sf T}_{\rm gc}^{-1}f\rangle$ and $F' \equiv \langle {\sf T'}_{\rm gc}^{-1}f'\rangle^{\prime}$  are defined as the gyroangle-averaged guiding-center push-forward transformations of the test-particle and field-particle distributions $f$ and $f'$, while the gyroangle-dependent parts $\wt{F} \equiv {\sf T}_{\rm gc}^{-1}f - \langle {\sf T}_{\rm gc}^{-1}f\rangle$ and $\wt{F}' \equiv {\sf T'}_{\rm gc}^{-1}f' - \langle {\sf T'}_{\rm gc}^{-1}f'\rangle^{\prime}$ are associated with higher-order (nonlinear) collisional effects that are usually neglected \cite{Brizard_2004,Sugama:2015}. 

By making use of the properties of the push-forward Lie-transform operator ${\sf T}_{\rm gc}^{-1}$: ${\sf T}_{\rm gc}^{-1}(f\,g) \equiv ({\sf T}_{\rm gc}^{-1}f)\,
({\sf T}_{\rm gc}^{-1}g)$, we find the following explicit expression for the two-particle guiding-center Landau tensor $\mathbb{Q}_{\rm gc}$:
\begin{equation}
\mathbb{Q}_{\rm gc} \equiv Q_{0}\,\delta^{3}({\bf x}_{\rm gc}^{\prime} - {\bf x}_{\rm gc})\;\left( \frac{\mathbb{I}}{|{\bf W}_{\rm gc}|} - 
\frac{{\bf W}_{\rm gc}\,{\bf W}_{\rm gc}}{|{\bf W}_{\rm gc}|^{3}} \right),
\label{eq:Q_gc_exact}
\end{equation}
where the guiding-center push-forward of the relative particle velocity ${\bf W}_{\rm gc} \equiv {\sf T}_{\rm gc}^{\prime -1}{\sf T}_{\rm gc}^{-1}{\bf w} $ is expressed as
\begin{eqnarray}
{\bf W}_{\rm gc} &\equiv& \frac{d_{\rm gc}{\bf x}_{\rm gc}}{dt} \;-\; \frac{d^{\prime}_{\rm gc}{\bf x}_{\rm gc}^{\prime}}{dt}  \;=\; {\bf W} \;+\; \epsilon \left(\frac{d_{\rm gc}\vb{\rho}_{\rm gc}}{dt} \;-\; \frac{d^{\prime}_{\rm gc}\vb{\rho}_{\rm gc}^{\prime}}{dt} \right),
\end{eqnarray}
where ${\bf W} \equiv d_{\rm gc}{\bf X}/dt - d^{\prime}_{\rm gc}{\bf X}^{\prime}/dt$ represents the relative guiding-center velocity, which is explicitly independent of the particle gyromotion, and the remaining finite Larmor-radius (FLR) terms include the relative polarization-drift velocity 
$d_{\rm gc}\vb{\rho}_{\rm E}/dt - d_{\rm gc}^{\prime}\vb{\rho}_{\rm E}^{\prime}/dt$ and the relative gyroangle-dependent perpendicular velocity $ d_{\rm gc}\vb{\rho}_{0}/dt - d^{\prime}_{\rm gc}\vb{\rho}_{0}^{\prime}/dt$.

\subsection{Guiding-center phase-space projection}

The guiding-center Fokker-Planck friction and diffusion coefficients in Eq.~\eqref{eq:C_gc} are expressed in Landau form as
\begin{eqnarray}
K_{\rm gc}^{\alpha}[F^{\prime}] &\equiv& \left\langle{\bf K}_{\rm gc}[F']\bdot\vb{\Delta}_{\rm gc}^{\alpha}\right\rangle  \;=\; \int_{\bf Z^{\prime}} B_{\|}^{\prime *} \left\langle \langle\mathbb{Q}_{\rm gc}\bdot\vb{\Delta}_{\rm gc}^{\alpha}\rangle\bdot
\vb{\Delta}_{\rm gc}^{\prime\beta}\right\rangle^{\prime} \pd{F^{\prime}}{Z'^{\beta}},   \label{eq:Kgc_Delta} \\
D_{\rm gc}^{\alpha\beta}[F'] &\equiv& \left\langle\vb{\Delta}_{\rm gc}^{\alpha}\bdot\mathbb{D}_{\rm gc}[F']\bdot\vb{\Delta}_{\rm gc}^{\beta}\right\rangle  \;=\; \int_{\bf Z'} B_{\|}^{\prime *} \left\langle \langle\vb{\Delta}_{\rm gc}^{\alpha}\bdot\mathbb{Q}_{\rm gc}\bdot\vb{\Delta}_{\rm gc}^{\beta}\rangle\right\rangle^{\prime} F^{\prime},
 \label{eq:Dgc_Delta}
\end{eqnarray}
where the guiding-center Fokker-Planck coefficients are ${\bf K}_{\rm gc}[F'] \equiv {\sf T}_{\rm gc}^{-1}{\bf K}[{\sf T}_{\rm gc}^{\prime}F']$ and $\mathbb{D}_{\rm gc}[F'] \equiv {\sf T}_{\rm gc}^{-1}\mathbb{D}[{\sf T}_{\rm gc}^{\prime}F']$, while 
$\mathbb{Q}_{\rm gc} \equiv {\sf T}_{\rm gc}^{\prime -1}{\sf T}_{\rm gc}^{-1}\mathbb{Q}$ denotes the guiding-center two-particle Landau tensor. In addition, the guiding-center Vlasov distribution $F' \equiv \langle {\sf T}_{\rm gc}^{\prime -1}f'\rangle^{\prime}$ is defined as the gyroangle-averaged guiding-center push-forward transformation of the Vlasov field-particle distribution $f'$. Here, the guiding-center projection vectors in reduced guiding-center test-particle phase space $({\bf X},p_{\|};\mu)$:
\begin{equation}
\vb{\Delta}_{\rm gc}^{\alpha} \;\equiv\; \frac{\Omega}{B}\,\pd{\vb{\rho}_{0}}{\zeta}\;\pd{Z^{\alpha}}{\mu} \;+\; \left\{ {\bf X},\frac{}{} Z^{\alpha}\right\}_{\rm rgc} 
\label{eq:Delta}
\end{equation}
is expressed in terms of the reduced guiding-center Poisson bracket \eqref{eq:PB_rgc}, where the reduced guiding-center phase-space components
\begin{eqnarray}
\vb{\Delta}_{\rm gc}^{\bf X} &=& \left\{ {\bf X},\frac{}{} {\bf X}\right\}_{\rm rgc} \;=\; \frac{\epsilon\,c\bhat}{qB_{\|}^{*}}\btimes\mathbb{I}, \label{eq:Delta_X} \\
\vb{\Delta}_{\rm gc}^{p_{\|}} &=& \left\{ {\bf X},\frac{}{} p_{\|}\right\}_{\rm rgc} \;=\; \frac{{\bf B}^{*}}{B_{\|}^{*}} \;=\; \frac{{\bf B}}{B_{\|}^{*}} \;+\; \epsilon\;\frac{p_{\|}c}{qB_{\|}^{*}}\;\nabla\btimes\bhat, \label{eq:Delta_p}
\end{eqnarray}
are separated from the magnetic-moment component
\begin{equation}
\vb{\Delta}_{\rm gc}^{\mu} \;=\; \frac{\Omega}{B}\,\pd{\vb{\rho}_{0}}{\zeta}.
\label{eq:Delta_mu}
\end{equation}
These projection-vector components satisfy the phase-space divergence property
\begin{equation}
\vb{\Delta}_{\rm gc}^{\alpha}\;\pd{F}{Z^{\alpha}} \;\equiv\; \frac{1}{B_{\|}^{*}}\;\pd{}{Z^{\alpha}}\left(\vb{\Delta}_{\rm gc}^{\alpha}\frac{}{} F_{\rm gc}\right),
\label{eq:Delta_div}
\end{equation}
which follows from the divergence property \eqref{eq:PB_gc_div} of the guiding-center Poisson bracket. We note that, while the gyromotion time scale has been eliminated from the guiding-center collision operator \eqref{eq:C_gc}, the guiding-center transformation has explicitly introduced spatial guiding-center friction and diffusion Fokker-Planck coefficients since $\vb{\Delta}_{\rm gc}^{\bf X} \neq 0$.

\subsection{Guiding-center magnetic-moment projection}

Collisional effects associated with the guiding-center magnetic moment $\mu$ involve the guiding-center projection vector \eqref{eq:Delta_mu}, which is explicitly gyroangle-dependent. Hence, collisional transport in $\mu$-space involves finite-gyroradius dynamics on the fast gyromotion time scale, which necessitates gyroangle averaging in Eqs.~\eqref{eq:C_gc}-\eqref{eq:Dgc_Delta}.

While the magnetic moment is an adiabatic invariant of the guiding-center dynamics, it is not an invariant of the exact particle dynamics. Hence, the classical diffusion process\cite{Hazeltine_Meiss} associated with the guiding-center projection vector $\vb{\Delta}_{\rm gc}^{\mu}$ is different from the nonadiabatic diffusion process associated with the breaking of the adiabatic invariance of the magnetic moment\cite{Bernstein_Rowlands_1976}, which occurs when a particle goes through an extremum point along a nonuniform magnetic-field line (where $|p_{\|}|$ reaches a maximum value) and induces a random jump in the magnetic moment. When the guiding-center magnetic moment is conserved (excluding small perturbations at extremal points along the guiding-center orbit), this nonadiabatic diffusion process is negligible.

\section{\label{sec:gcL}Dissipative guiding-center Landau Bracket}

The metriplectic structure \cite{Hirvijoki_2022} for the time evolution of any guiding-center functional ${\cal F}[F_{\rm gc},{\bf D}_{\rm gc},{\bf B}]$ can be determined by the differential equation
\begin{equation} 
\pd{\cal F}{t} \;=\; [{\cal F},\mathcal{H}_{\rm gc}]_{\rm gc} \;+\; ({\cal F},\mathcal{S}_{\rm gc})_{\rm gc},
\label{eq:gc_metri}
\end{equation}
where the symmetric guiding-center dissipative bracket $(\cdot,\cdot)_{\rm gc}$ must satisfy the following conservation properties $({\cal E}_{\rm gc}^{a},{\cal G})_{\rm gc} = 0$, where ${\cal E}_{\rm gc}^{a} = ({\cal H}_{\rm gc},{\cal P}_{\rm gc}^{i},
{\cal P}_{{\rm gc}\varphi})$ are Hamiltonian invariants and ${\cal G}$ is an arbitrary functional (hence, these invariants may be viewed as dissipative Casimir invariants), and an H-theorem $(\mathcal{S}_{\rm gc},\mathcal{S}_{\rm gc})_{\rm gc} \geq 0$. 

\subsection{Guiding-center Landau collision operator}

Following our previous work \cite{Hirvijoki_2022}, we define the guiding-center two-particle vector field 
\begin{equation}
\vb{\Gamma}_{\rm gc}[\mathcal{A}]({\bf Z};{\bf Z}^{\prime}) \equiv \vb{\gamma}_{\rm gc}[\mathcal{A}]({\bf Z}) - \vb{\gamma}_{\rm gc}[\mathcal{A}]({\bf Z}^{\prime}), 
\label{eq:Gamma_A_def}
\end{equation}
where the test-particle contribution is defined as
\begin{equation}
\vb{\gamma}_{\rm gc}[\mathcal{A}]({\bf Z}) \equiv \frac{\Omega}{B}\,\pd{\vb{\rho}_{0}}{\zeta}\;\pd{}{\mu}\left(\fd{\cal A}{F_{\rm gc}({\bf Z})}\right) + \left\{{\bf X},\; \fd{\cal A}{F_{\rm gc}({\bf Z})}\right\}_{\rm rgc} + 4\pi\,q\;\frac{\delta^{\star}{\cal A}}{\delta{\bf D}_{\rm gc}({\bf X})}\bdot
\left\{{\bf X},\; {\bf X}\right\}_{\rm rgc},
 \label{eq:gamma_rgc}
 \end{equation}
which combines the guiding-center transformation of the particle Poisson bracket appearing in Eq.~\eqref{eq:gamma_particle}:
\[ {\sf T}_{\rm gc}^{-1}\left\{{\bf x},\; \fd{\cal A}{f}\right\} \;\equiv\; \left\{ {\sf T}_{\rm gc}^{-1}{\bf x},\; \fd{\cal A}{F_{\rm gc}}\right\}_{\rm gc} \;\simeq\; \frac{\Omega}{B}\,\pd{\vb{\rho}_{0}}{\zeta}\;\pd{}{\mu}\left(\fd{\cal A}{F_{\rm gc}({\bf Z})}\right) + \left\{{\bf X},\; \fd{\cal A}{F_{\rm gc}({\bf Z})}\right\}_{\rm rgc} \]
with the reduced guiding-center displacement term\cite{Hirvijoki_2022} $4\pi q\,\delta^{\star}{\cal A}/\delta{\bf D}_{\rm gc}\bdot\{{\bf X},\,{\bf X}\}_{\rm rgc}$, where the drift-kinetic approximation ${\sf T}_{\rm gc}^{-1}{\bf x} = {\bf X} + \epsilon\,\vb{\rho}_{\rm gc} \simeq {\bf X}$ is used in the reduced guiding-center Poisson bracket $\{{\sf T}_{\rm gc}^{-1}{\bf x},\, \delta{\cal A}/\delta F_{\rm gc}\}_{\rm rgc} \simeq \{{\bf X},\; \delta{\cal A}/\delta F_{\rm gc}\}_{\rm rgc}$. This definition yields the particle guiding-center velocity
\begin{eqnarray} 
\vb{\gamma}_{\rm gc}[{\cal H}_{\rm gc}]({\bf Z}) &=&  \Omega\,\pd{\vb{\rho}_{0}}{\zeta} \;+\; \frac{p_{\|}}{m}\,\frac{{\bf B}^{*}}{B_{\|}^{*}} \;+\; \left(q\,{\bf E}^{\star} \;-\frac{}{} \epsilon\,\nabla K_{\rm gc}\right)\btimes\frac{c\bhat}{qB_{\|}^{*}} \nonumber \\
 &=& \Omega\,\pd{\vb{\rho}_{0}}{\zeta} \;+\; \frac{d_{\rm gc}{\bf X}}{dt} \;\equiv\; {\bf v}_{\rm gc},
 \end{eqnarray}
where only the lowest order contribution $d_{\rm gc}\vb{\rho}_{\rm gc}/dt = \Omega\,\partial\vb{\rho}_{0}/\partial\zeta + \cdots$ is retained. Hence, the relative guiding-center velocity is
\begin{equation} 
\vb{\Gamma}_{\rm gc}[{\cal H}_{\rm gc}]({\bf Z};{\bf Z}') \;=\;  {\bf v}_{\rm gc} -  {\bf v}_{\rm gc}^{\prime} \equiv  \left(\Omega\,\pd{\vb{\rho}_{0}}{\zeta} -  \Omega'\,\pd{\vb{\rho}^{\prime}_{0}}{\zeta'}\right) \;+\; \vb{\Gamma}_{\rm rgc}[{\cal H}_{\rm gc}] \;=\; {\bf w}_{\rm gc},
\label{eq:Gamma_W}
\end{equation}
and the symmetric guiding-center two-particle Landau tensor \eqref{eq:Q_gc_exact} is henceforth approximated  as
\begin{equation}
\mathbb{Q}_{\rm gc} \;\equiv\; Q_{0}\,\delta^{3}({\bf X}^{\prime} - {\bf X})\;\left( \frac{\mathbb{I}}{|{\bf w}_{\rm gc}|} - \frac{{\bf w}_{\rm gc}{\bf w}_{\rm gc}}{|{\bf w}_{\rm gc}|^{3}} \right),
\label{eq:Q_gc}
\end{equation}
which is now evaluated locally in guiding-center space (i.e., FLR corrections are ignored in the spatial delta function). The guiding-center dissipative Landau bracket is defined as
\begin{eqnarray}
(\mathcal{A},\mathcal{B})_{\rm gc} &=& \frac{1}{2}\int_{Z, Z'} \vb{\Gamma}_{\rm gc}[\mathcal{A}]\bdot \mathbb{Q}_{\rm gc} \bdot \vb{\Gamma}_{\rm gc}[\mathcal{B}]  \nonumber \\
 &=& \frac{Q_{0}}{2}\int_{Z, Z'}\frac{\delta^{3}({\bf X}^{\prime} - {\bf X})}{|{\bf w}_{\rm gc}|^{3}}\;{\bf w}_{\rm gc}\btimes\vb{\Gamma}_{\rm gc}[\mathcal{A}]\bdot {\bf w}_{\rm gc}\btimes\vb{\Gamma}_{\rm gc}[\mathcal{B}],
\label{eq:gc_diss_br}
\end{eqnarray}
where we use the notation
\[ \int_{Z, Z'} (\cdots) \;=\; \sum\int_{{\bf Z}}F_{\rm gc}({\bf Z})\,\left\langle \sum^{\prime}\int_{{\bf Z'}}F_{\rm gc}^{\prime}({\bf Z'})\frac{}{}\langle\cdots\rangle^{\prime}\right\rangle. \]
We note that the energy conservation property $({\cal H}_{\rm gc},{\cal B})_{\rm gc} = 0$ of the guiding-center dissipative Landau bracket \eqref{eq:gc_diss_br} is automatic (for all ${\cal B}$), since the two-particle vector \eqref{eq:Gamma_W} is a null-vector of the guiding-center Landau tensor \eqref{eq:Q_gc}: ${\bf w}_{\rm gc}\bdot\mathbb{Q}_{\rm gc} \;\equiv\; 0$.

Using Eq.~\eqref{eq:P_gc_i}, the conservation of the guiding-center momentum $({\cal P}_{\rm gc}^{i},{\cal G})_{\rm gc} = 0$ follows from the expression
\begin{equation}
\vb{\Gamma}_{\rm gc}[{\cal P}_{\rm gc}^{i}]({\bf Z};{\bf Z'}) \;=\; \wh{\imath} \;-\; \wh{\imath} \;\equiv\; 0, 
\end{equation}
where we find
\begin{eqnarray}
\vb{\gamma}_{\rm gc}[{\cal P}_{\rm gc}^{i}]({\bf Z}) &=& \frac{{\bf B}^{*}}{B_{\|}^{*}}\;b^{i} \;+\; \left[ \frac{q}{c}\,{\bf B}\btimes\wh{\imath} \;+\; \epsilon\,\frac{p_{\|}}{B}\;\nabla\btimes\left({\bf B}\btimes\wh{\imath}\right) \;-\; \epsilon\,p_{\|}\;\nabla b^{i} \right]\btimes\frac{c\bhat}{qB_{\|}^{*}} \nonumber \\
  &=& \wh{\imath}\;\frac{B}{B_{\|}^{*}} \;+\; \epsilon\,\frac{p_{\|}c}{qB_{\|}^{*}} \left[\bhat\btimes\nabla b^{i} \;+\;  (\nabla\btimes\bhat)\,b^{i} \;+\; \nabla\btimes({\bf B}\btimes\wh{\imath})\btimes \frac{\bhat}{B} \right] \nonumber \\
    &=& \wh{\imath}\;\frac{B}{B_{\|}^{*}} \;+\; \epsilon\,\frac{p_{\|}c}{qB_{\|}^{*}} \left[\bhat\btimes\nabla b^{i} \;+\; \left(\nabla\btimes\bhat\right)\,b^{i} \;+\;  \left( \nabla b^{i} \;-\; \wh{\imath}\btimes(\nabla\btimes\bhat)\right) \btimes\bhat\right] \nonumber \\
 &=& \frac{\wh{\imath}}{B_{\|}^{*}} \left( B \;+\; \epsilon\,\frac{p_{\|}c}{q}\,\bhat\bdot\nabla\btimes\bhat\right) \;=\; \wh{\imath}.
\end{eqnarray}
Using the same identity, Eq.~\eqref{eq:P_gc_phi} yields the expression
\begin{equation} 
\vb{\Gamma}_{\rm gc}[{\cal P}_{{\rm gy}\varphi}]({\bf Z};{\bf Z}') \;=\; \pd{{\bf X}}{\varphi} - \pd{{\bf X}^{\prime}}{\varphi'} \;=\; \wh{\sf z}\btimes\left({\bf X} \;-\frac{}{} {\bf X}^{\prime}\right),
\end{equation}
and the conservation of guiding-center toroidal canonical angular momentum $({\cal P}_{{\rm gc}\varphi},{\cal G})_{\rm gc} = 0$ follows from the fact that $\mathbb{Q}_{\rm gc}$ contains the spatial delta function 
$\delta^{3}({\bf X}^{\prime} - {\bf X})$. Since the three functionals $({\cal H}_{\rm gc}, {\cal P}_{\rm gc}^{i}, {\cal P}_{{\rm gc}\varphi})$ are conserved by the guiding-center dissipative Landau bracket \eqref{eq:gc_diss_br} for any functional ${\cal G}$, they are called guiding-center collisional Casimir invariants. We note that the interior guiding-center Casimir functional ${\cal M}_{\rm gc}$ associated with the adiabatic invariance of the magnetic moment $\mu$ yields the test-particle contribution
\begin{equation}
\vb{\gamma}_{\rm gc}[{\cal M}_{\rm gc}] \;=\; \frac{\Omega}{B}\,\pd{\vb{\rho}_{0}}{\zeta} \;\equiv\; \vb{\Delta}_{\rm gc}^{\mu}({\bf X},\mu,\zeta),
\end{equation}
which implies that, in our work, ${\cal M}_{\rm gc}$ is not a guiding-center collisional Casimir invariant and, therefore, our work describes classical transport processes associated with $\vb{\Delta}_{\rm gc}^{\mu} \neq 0$ as well as classical spatial diffusion associated with $\vb{\Delta}_{\rm gc}^{\bf X}$ (see App.~\ref{sec:Isotropic}). Here, the collisional evolution of the guiding-center magnetic-moment functional ${\cal M}_{\rm gc}$ is expressed as
\begin{eqnarray}
\pd{{\cal M}_{\rm gc}}{t} &\equiv& \left({\cal M}_{\rm gc},\frac{}{} {\cal S}_{\rm gc} \right)_{\rm gc} \;=\; \frac{Q_{0}}{2} \int_{Z,Z'} \frac{\delta^{3}({\bf X}' - {\bf X})}{|{\bf w}_{\rm gc}|^{3}}\;{\bf w}_{\rm gc}\btimes\vb{\Gamma}_{\rm gc}[{\cal M}_{\rm gc}]\bdot {\bf w}_{\rm gc}\btimes \vb{\Gamma}_{\rm gc}[{\cal S}_{\rm gc}] \nonumber \\
 &=&\sum\int_{\bf Z} F_{\rm gc}\,\left[ \sum^{\prime}\int_{\bf Z'}\;F^{\prime}_{\rm gc}\;\left\langle \vb{\Delta}_{\rm gc}^{\mu}\bdot\frac{}{}\langle \mathbb{Q}_{\rm gc}\bdot\vb{\Gamma}_{\rm gc}[{\cal S}_{\rm gc}]\rangle^{\prime} \right\rangle\right] \nonumber \\
  &\equiv& \sum\int_{\bf Z} B_{\|}^{*} \left( F\;K_{\rm gc}^{\mu} \;-\; D_{\rm gc}^{\mu\beta}\;\pd{F}{Z^{\beta}} \right),
 \label{eq:M_FP}
\end{eqnarray}
where $\vb{\Delta}_{\rm gc}^{\mu}$ contributes to the guiding-center magnetic-moment friction $K_{\rm gc}^{\mu}$ and the guiding-center magnetic-moment diffusion components $(D_{\rm gc}^{\mu{\bf X}},D_{\rm gc}^{\mu p_{\|}},D_{\rm gc}^{\mu\mu})$.

In their recent work, Sato and Morrison\cite{Sato_Morrison_2024,Sato_Morrison_2025} use an approximation of weak collisionality in which the collisional mean-free-path is much larger than the magnetic-gradient length scale (which is itself much larger than the gyroradii of all particles). Based on this neoclassical approximation\cite{Hazeltine_Meiss}, both test and field particles have enough time to complete their respective guiding-center orbits (for both trapped and circulating particles) before they experience a collision at a common guiding-center position. These full guiding-center orbits, however, assume that their respective magnetic moments are conserved, which allows Sato and Morrison\cite{Sato_Morrison_2024,Sato_Morrison_2025} to assume that ${\cal M}_{\rm gc}$ should be considered a guiding-center collisional Casimir invariant under this weak-collisionality approximation. This approximation is implicitly consistent with removing the gyromotion component from the guiding-center Poisson bracket \eqref{eq:PB_gc}, so that this reduced guiding-center Poisson bracket in reduced guiding-center space $({\bf X},p_{\|};\mu)$ yields $\{\mu,\;g\}_{\rm rgc} \equiv 0$, and only the projection components $(\vb{\Delta}_{\rm gc}^{\bf X},\vb{\Delta}_{\rm gc}^{p_{\|}})$ remain in the reduced guiding-center Fokker-Planck collision operator, without compromising the guiding-center collisional invariants (which now includes the guiding-center magnetic-moment functional ${\cal M}_{\rm gc}$). We, therefore, note that our work includes the work of Sato and Morrison\cite{Sato_Morrison_2024,Sato_Morrison_2025} (regarding its guiding-center formulation) by setting $\vb{\Delta}_{\rm gc}^{\mu} = 0$ and using the relative two-particle guiding-center velocity ${\bf W} = \vb{\Gamma}_{\rm rgc}[{\cal H}_{\rm gc}]$ in the guiding-center two-particle Landau tensor \eqref{eq:Q_gc}.

Next, the guiding-center dissipative Landau bracket \eqref{eq:gc_diss_br} satisfies an H-theorem:
\begin{equation}
({\cal S}_{\rm gc},{\cal S}_{\rm gc})_{\rm gc} \;=\;\frac{Q_{0}}{2}\int_{Z, Z'}\delta^{3}({\bf X}^{\prime} - {\bf X})\frac{\left|{\bf w}_{\rm gc}\btimes\vb{\Gamma}_{\rm gc}[{\cal S}_{\rm gc}]\right|^{2}}{|{\bf w}_{\rm gc}|^{3}} \geq 0,
\label{eq:gc_H}
\end{equation}
where
\begin{equation}
\vb{\Gamma}_{\rm gc}[{\cal S}_{\rm gc}] \;=\; \vb{\gamma}_{\rm gc}[{\cal S}_{\rm gc}]({\bf Z})  \;-\;  \vb{\gamma}_{\rm gc}[{\cal S}_{\rm gc}]({\bf Z}^{\prime}).
\end{equation}
Here, the test-particle guiding-center term $\vb{\gamma}_{\rm gc}[{\cal S}_{\rm gc}]({\bf Z})$ is expressed as
\begin{equation}
 \vb{\gamma}_{\rm gc}[{\cal S}_{\rm gc}]({\bf Z}) \;=\; -\,\frac{\Omega}{B}\,\pd{\vb{\rho}_{0}}{\zeta}\,\pd{\ln F}{\mu} \;-\; \left\{ {\bf X},\frac{}{} \ln F\right\}_{\rm rgc}  \;\equiv\; -\;\frac{1}{F_{\rm gc}}\;\pd{}{Z^{\alpha}}\left( F_{\rm gc}\;\vb{\Delta}_{\rm gc}^{\alpha}\right),
   \label{eq:gamma_Sgc}
 \end{equation}
where we used the identity $\partial(B_{\|}^{*}\vb{\Delta}_{\rm gc}^{\alpha})/\partial Z^{\alpha} \equiv 0$. The guiding-center thermodynamic equilibrium $({\cal S}_{\rm gc},{\cal S}_{\rm gc})_{\rm gc} = 0$, therefore, requires that ${\bf w}_{\rm gc}\btimes
\vb{\Gamma}_{\rm gc}[{\cal S}_{\rm gc}] = 0$. If we substitute the guiding-center Maxwell-Boltzmann  distribution $F_{\rm MBgc} = F_{0}({\bf X})\;\exp[- H_{\rm gc}/T({\bf X})]$ into Eq.~\eqref{eq:gamma_Sgc}, where $H_{\rm gc} \equiv \epsilon^{-1}q\,\Phi + K_{\rm gc}$, we find
\[ -\,\frac{\Omega}{B}\,\pd{\vb{\rho}_{0}}{\zeta}\,\pd{\ln F}{\mu} \;-\; \frac{{\bf B}^{*}}{B_{\|}^{*}}\,\pd{\ln F }{p_{\|}} \;-\; \frac{\epsilon\,c\bhat}{qB_{\|}^{*}}\btimes\nabla \ln F \;=\; \frac{{\bf v}_{\rm gc}}{T} \;-\; \frac{\epsilon\,c\bhat}{qB_{\|}^{*}}\btimes\nabla\left( \ln F_{0}  \;+\; \frac{H_{\rm gc}}{T}\;
\nabla \ln T\right), \]
and, as expected, thermodynamic equilibrium requires equal temperatures for all particle species and a spatially uniform plasma. We note that, by including the total guiding-center magnetic-moment functional ${\cal M}_{\rm gc}$ as a collisional invariant, Sato and Morrison\cite{Sato_Morrison_2024,Sato_Morrison_2025} increase the class of guiding-center equilibria by allowing arbitrary functions of the guiding-center magnetic moment $\mu$, which allow for arbitrary deviations from the guiding-center Maxwell-Boltzmann  distribution $F_{\rm MBgc}$.

Lastly, in what follows, we will introduce the notation
 \begin{eqnarray}
 {\bf S}_{\rm gc}[F']({\bf Z}) &\equiv&  \int_{Z^{\prime}} \mathbb{Q}_{\rm gc}\bdot\left( \frac{1}{F^{\prime}}\; \vb{\Delta}_{\rm gc}^{\prime\,\beta}\;\pd{F^{\prime}}{Z^{\prime\beta}} - \frac{1}{F}\; \vb{\Delta}_{\rm gc}^{\beta}\;\pd{F}{Z^{\beta}} \right) \nonumber \\
  &\equiv& {\bf K}_{\rm gc}[F^{\prime}]({\bf Z}) \;-\; \mathbb{D}_{\rm gc}[F^{\prime}]({\bf Z})\bdot \frac{\vb{\Delta}_{\rm gc}^{\beta}}{F}\;\pd{F}{Z^{\beta}},
 \label{eq:Sgc_flux}
 \end{eqnarray}
 which defines the guiding-center collisional flux. We note that it is the guiding-center equivalent of the collisional flux defined in Eq.~\eqref{eq:C_FP} and it obeys the same symmetry property as in Eq.~\eqref{eq:FPL_symmetry}:
 \begin{equation}
\int_{\bf Z}F_{\rm gc}({\bf Z})\;\delta^{3}({\bf X} - {\bf X}_{0})\,{\bf S}_{\rm gc}[F']({\bf Z}) \;\equiv\; -\;\int_{{\bf Z}'} F'_{\rm gc}({\bf Z}')\;\delta^{3}({\bf X}' - {\bf X}_{0})\,{\bf S}_{\rm gc}[F]({\bf Z}'),
\label{eq:FPLgc_symmetry}
\end{equation}
where ${\bf X}_{0}$ denotes an arbitrary field point and, consistent with the local approximation used in Eq.~\eqref{eq:Q_gc}, FLR corrections have been omitted in the spatial delta function. The symmetry property \eqref{eq:FPLgc_symmetry}, therefore, guarantees that the local guiding-center collisional force 
\begin{equation}
{\bf F}_{\rm gc}^{\sf C}({\bf X}_{0}) \;\equiv\; \int_{P}F_{\rm gc}({\bf Z})\;\delta^{3}({\bf X} - {\bf X}_{0})\,\sum^{\prime}{\bf S}_{\rm gc}[F']({\bf Z}) 
\label{eq:Force_gc_coll} 
\end{equation}
satisfies the momentum-conservation property $\sum{\bf F}_{\rm gc}^{\sf C}({\bf X}_{0}) = 0$.

\subsection{Guiding-center Fokker-Planck collision operator}

Using the guiding-center metriplectic structure presented above, the time evolution of the guiding-center functional $\Psi_{V} \equiv \int_{\bf Z}\psi\,F_{\rm gc}$, where $\psi({\bf X},p_{\|},\mu)$ is an arbitrary function, is expressed in metriplectic form as
\begin{eqnarray}
\pd{\Psi_{V}}{t} &=& \int_{\bf Z}\psi\;\pd{F_{\rm gc}}{t} \;=\; \int_{\bf Z}\psi\; \left[ B_{\|}^{*}\,\pd{F}{t} - F\;\nabla\bdot\left(B_{\|}^{*}\frac{d_{\rm gc}{\bf X}}{dt}\right) - F\;\pd{}{p_{\|}}\left(B_{\|}^{*}\frac{d_{\rm gc}p_{\|}}{dt}\right)\right] \\
 &=& \int_{\bf Z} \left[\psi\,B_{\|}^{*}\;\frac{d_{\rm gc}F}{dt} + F_{\rm gc} \left( \{\psi,\; K_{\rm gc}\}_{\rm rgc} \;+\frac{}{} \epsilon^{-1}q\,{\bf E}^{\star}\bdot\{{\bf X},\; \psi\}_{\rm rgc} \right) \right] \nonumber \\
  &\equiv& (\Psi_{V},{\cal S}_{\rm gc})_{\rm gc} \;+\; [\Psi_{V},\; {\cal H}_{\rm gc}]_{\rm gc}, \nonumber 
\end{eqnarray}
where the guiding-center Hamiltonian bracket is
\[ [\Psi_{V},\; {\cal H}_{\rm gc}]_{\rm gc}  \;=\; \int_{\bf Z} F_{\rm gc} \left( \{\psi,\; K_{\rm gc}\}_{\rm rgc} \;+\frac{}{} \epsilon^{-1}q\,{\bf E}^{\star}\bdot\{{\bf X},\; \psi\}_{\rm rgc} \right), \]
while the guiding-center Landau collision operator 
\begin{eqnarray} 
(\Psi_{V},{\cal S}_{\rm gc})_{\rm gc}  &=& \int_{\bf Z} \left[\left(\frac{\Omega}{B}\,\pd{\vb{\rho}_{0}}{\zeta}\;\pd{\psi}{\mu} + \frac{{\bf B}^{*}}{B_{\|}^{*}}\pd{\psi}{p_{\|}} \right) - \epsilon\,\nabla\psi\btimes\frac{c\bhat}{qB_{\|}^{*}}\right]\bdot \sum^{\prime}F_{\rm gc}\,{\bf S}_{\rm gc}[F'] 
\nonumber \\
 &=& \int_{\bf Z} \left(\vb{\Delta}_{\rm gc}^{\alpha}\;\pd{\psi}{Z^{\alpha}}\right)\vb\bdot \sum^{\prime}F_{\rm gc}\,{\bf S}_{\rm gc}[F']  \;\equiv\; \int_{\bf Z} B_{\|}^{*}\psi\;\sum^{\prime}{\cal C}_{\rm gc}[F; F']  
 \label{eq:gcLandau_FP}
\end{eqnarray}
is now expressed in Fokker-Planck form as
\begin{eqnarray}
{\cal C}_{\rm gc}[F; F'] &=& -\;\frac{1}{B_{\|}^{*}}\pd{}{Z^{\alpha}}\left(B_{\|}^{*}\vb{\Delta}_{\rm gc}^{\alpha}\bdot F\,{\bf S}_{\rm gc}[F']\right) \nonumber \\
 &\equiv& -\; \frac{1}{B_{\|}^{*}}\pd{}{Z^{\alpha}}\left[ B_{\|}^{*}\vb{\Delta}_{\rm gc}^{\alpha}\bdot\left( {\bf K}_{\rm gc}[F']\,F - \mathbb{D}_{\rm gc}[F']\bdot \vb{\Delta}_{\rm gc}^{\beta}\pd{F}{Z^{\beta}}\right)\right],
\label{eq:C_gc_FP} 
\end{eqnarray}
where ${\bf K}_{\rm gc}[F']$ and $\mathbb{D}_{\rm gc}[F']$ are defined in Eq.~\eqref{eq:Sgc_flux} as functionals of the field-particle guiding-center Vlasov distribution $F'$. In Eq.~\eqref{eq:C_gc_FP}, and henceforth, gyroangle-averaging is implied wherever appropriate since the test-particle guiding-center distribution $F$ is gyroangle-independent. 

Lastly, we note that the time evolution of an arbitrary moment of the guiding-center Vlasov distribution $F_{\rm gc}$ is expressed as 
\begin{eqnarray}
\pd{}{t}\left(\int_{P}\chi F_{\rm gc}\right) &=& -\; \nabla\bdot\left[\int_{P}\chi F_{\rm gc}\left( \frac{d_{\rm gc}{\bf X}}{dt} - \frac{\epsilon\,c\bhat}{q B_{\|}^{*}}\btimes\sum^{\prime}{\bf S}_{\rm gc}\right)\right] \nonumber \\
 &&+\; \int_{P} F_{\rm gc}\left( \frac{d_{\rm gc}\chi}{dt} + \sum^{\prime}{\bf S}_{\rm gc}\bdot\vb{\Delta}_{\rm gc}^{\alpha}\pd{\chi}{Z^{\alpha}}\right),
\label{eq:chi_moment}
\end{eqnarray} 
where $\chi({\bf X},p_{\|},\mu)$ is an arbitrary function on guiding-center phase space. Important applications of this guiding-center moment equation include the guiding-center particle number $(\chi = 1)$ conservation law for each charged-particle species
\begin{equation}
\pd{N_{\rm gc}}{t} \;+\; \nabla\bdot\left(N_{\rm gc}\frac{}{}{\bf U}_{\rm gc} \right) \;=\; \nabla\bdot\left(\int_{P}F_{\rm gc}\,\frac{\epsilon c\bhat}{qB_{\|}^{*}}\btimes\sum^{\prime}\langle{\bf S}_{\rm gc}\rangle\right) \;\simeq\; \nabla\bdot\left(\frac{\epsilon c\bhat}{qB}\btimes{\bf F}_{\rm gc}^{\sf C}\right),
\end{equation}
where we used the approximation $B_{\|}^{*} \simeq B$ and the definition \eqref{eq:Force_gc_coll} for the local guiding-center collisional force, while gyroangle-averaging of the guiding-center collisional flux \eqref{eq:Sgc_flux} is now shown explicitly. If we define the guiding-center collisional fluid flow
\begin{equation}
N_{\rm gc}{\bf U}_{\rm gc}^{\sf C} \;\equiv\; -\,\frac{\epsilon c\bhat}{qB}\btimes{\bf F}_{\rm gc}^{\sf C}, 
\label{eq:coll_gcflow}
\end{equation}
then the momentum-conservation property $\sum{\bf F}_{\rm gc}^{\sf C} = 0$ of the guiding-center collisional force automatically guarantees the ambipolarity condition\cite{Hazeltine_Meiss} on the collisional guiding-center current density:
\begin{equation}
{\bf J}_{\rm gc}^{\sf C} \;\equiv\; \sum \epsilon^{-1}qN_{\rm gc}{\bf U}_{\rm gc}^{\sf C} \;=\; -\;\frac{c\bhat}{B}\btimes\left(\sum{\bf F}_{\rm gc}^{\sf C}\right) \;=\; 0.
\label{eq:ambipolar}
\end{equation} 
Similar moments of the guiding-center Fokker-Planck collision operator has appeared elsewhere \cite{Sugama_2017,Sugama_2017_RMPP}. In App.~\ref{sec:Isotropic}, we will present a simple model that shows that guiding-center collisional fluid flow \eqref{eq:coll_gcflow} is diffusive in character\cite{Hazeltine_Meiss}.

\subsection{Guiding-center Amp\`{e}re-Maxwell-Landau equation}

 Since the guiding-center two-particle vector field $\vb{\Gamma}_{\rm gc}[\mathcal{A}]({\bf Z};{\bf Z}')$, defined in Eq.~\eqref{eq:Gamma_A_def}, involves a functional derivative of a general functional ${\cal A}$ with respect to the guiding-center displacement field ${\bf D}_{\rm gc}$, then the guiding-center Amp\`{e}re-Maxwell acquires a dissipative collisional current. 
 
 Indeed, consider the functional $\Psi[{\bf D}_{\rm gc}] = \int_{X} \vb{\alpha}\bdot{\bf D}_{\rm gc}$, where $\vb{\alpha}({\bf X})$ is an arbitrary vector function, so that 
\begin{equation} 
\pd{\Psi}{t} \;=\; \int_{X} \vb{\alpha}\bdot\pd{{\bf D}_{\rm gc}}{t} \;=\; [\Psi,\mathcal{H}_{\rm gc}]_{\rm gc}+(\Psi,\mathcal{S}_{\rm gc})_{\rm gc},
\end{equation}
is expressed in terms of the Hamiltonian bracket
\begin{eqnarray*}
[\Psi, {\cal H}_{\rm gc}]_{\rm gc} &=& 4\pi c\int_{X} \vb{\alpha}\bdot\nabla\btimes\left(\frac{\bf B}{4\pi} \;-\; \sum\int_{P} F_{\rm gc}\;\vb{\mu}_{\rm gc}\right) \nonumber \\
 &&-\; 4\pi \sum \int_{Z} F_{\rm gc}\;\left(q\,\vb{\alpha} + q\,\mathbb{P}_{\|}\bdot\nabla\btimes\vb{\alpha}\right) \bdot\frac{d_{\rm gc}{\bf X}}{dt} \\
 &\equiv& \int_{X} \vb{\alpha}\bdot\left( c\,\nabla\btimes{\bf H}_{\rm gc} \;-\frac{}{} 4\pi\,{\bf J}_{\rm gc}\right),
 \end{eqnarray*}
where the guiding-center current density ${\bf J}_{\rm gc}$ defined in Eq.~\eqref{eq:Maxwell_eq} and ${\bf H}_{\rm gc} \equiv {\bf B} - 4\pi\,\vb{\sf M}_{\rm gc}$ is defined in terms of the guiding-center magnetization \eqref{eq:M_def}, while the dissipative bracket is expressed as
 \begin{eqnarray*}
(\Psi, {\cal S}_{\rm gc})_{\rm gc} &=& 4\pi c \sum\int_{\bf Z}\left(\vb{\alpha} \;+\; \epsilon\,\mathbb{P}_{\|}\bdot\nabla\btimes\vb{\alpha}\right)\bdot\frac{\bhat}{B_{\|}^{*}}\btimes \sum^{\prime}\nonumber \\
 &=& 4\pi c \sum\int_{\bf Z} \vb{\alpha}\bdot\sum^{\prime}\left[ \frac{\bhat}{B_{\|}^{*}}\btimes F_{\rm gc}\,\langle{\bf S}_{\rm gc}[F']\rangle \;-\; \nabla\btimes\left( \frac{q}{c}\,\mathbb{P}_{\|}\bdot\frac{\epsilon c\bhat}{qB_{\|}^{*}}\btimes F_{\rm gc}\,\langle{\bf S}_{\rm gc}[F']\rangle\right) \right]
 \nonumber \\
  &\equiv& -\;4\pi\int_{X} \vb{\alpha}\bdot\left(  {\bf J}_{\rm gc}^{\sf C} \;+\frac{}{} c\,\nabla\btimes\vb{\sf M}_{\rm gc}^{\sf C}\right),
\end{eqnarray*}
where the collisional guiding-center current density ${\bf J}_{\rm gc}^{\sf C}$ is shown to vanish as a result of the ambipolarity condition \eqref{eq:ambipolar}, while there is a nonvanishing collisional contribution to guiding-center magnetization.
Hence, the guiding-center Amp\`ere--Maxwell-Landau equation becomes
\begin{equation}  
\pd{{\bf D}_{\rm gc}}{t} \;=\; c\,\nabla\btimes{\bf H}_{\rm gc}^{\sf C} \;-\; 4\pi\,{\bf J}_{\rm gc},
 \label{eq:Ampere_Landau}
 \end{equation} 
which includes a collisional contribution to the guiding-center magnetization:
\begin{equation}
{\bf H}_{\rm gc}^{\sf C} \equiv {\bf B} \;-\; 4\pi \epsilon\,\int_{P} F_{\rm gc} \left[ \vb{\mu}_{\rm gc} \;+\; \frac{q}{c}\,\mathbb{P}_{\|}\bdot\left(\frac{d_{\rm gc}{\bf X}}{dt} \;-\; \frac{\epsilon c\bhat}{qB_{\|}^{*}}\btimes\sum^{\prime}\langle{\bf S}_{\rm gc}[F']\rangle \right) \right], 
\label{eq:Hgc_col} 
\end{equation}
Here, we see how the metriplectic bracket formulation allows for a self-consistent combination of the dissipationless and dissipative contributions to the guiding-center current density and the guiding-center moving electric-dipole contribution to the guiding-center magnetization.

\subsection{Drift-kinetic limit of the guiding-center Landau collision operator}

We conclude this Section by considering the drift-kinetic limit, in which we retain only the lowest-order terms in powers of the guiding-center ordering parameter $\epsilon$. In this limit, the guiding-center velocity becomes
\begin{equation}
{\bf v}_{\rm gc0} \;=\; \Omega\,\pd{\vb{\rho}_{0}}{\zeta} \;+\; \frac{p_{\|}}{m}\,\bhat \;+\; {\bf E}\btimes\frac{c\bhat}{B} \;\equiv\; {\bf v} \;+\; {\bf E}\btimes\frac{c\bhat}{B},
\label{eq:v_0}
\end{equation}
which is simply the sum of the particle velocity ${\bf v}$ and the $E\times B$ velocity, and the guiding-center Jacobian simply becomes $mB$. Since the latter velocity is the same for all particle species, the lowest-order relative guiding-center velocity 
${\bf w}_{\rm gc0} = {\bf w}$ is identical to the relative velocity \eqref{eq:u_def}.

In the drift-kinetic limit, the guiding-center two-particle vector field \eqref{eq:Gamma_A_def} becomes $\vb{\Gamma}_{\rm gc0}[{\cal A}]({\bf Z};{\bf Z}') \equiv \vb{\gamma}_{\rm gc0}[{\cal A}]({\bf Z}) - \vb{\gamma}_{\rm gc0}[{\cal A}]({\bf Z}')$, where
\begin{equation}
\vb{\gamma}_{\rm gc0}[{\cal A}]({\bf Z}) \;\equiv\; \frac{\Omega}{B}\,\pd{\vb{\rho}_{0}}{\zeta}\;\pd{}{\mu}\left(\fd{\cal A}{F_{\rm gc}}\right) \;+\; \bhat\;\pd{}{p_{\|}}\left(\fd{\cal A}{F_{\rm gc}}\right) \;+\; 4\pi c\;\fd{\cal A}{{\bf D}_{\rm gc}}\btimes\frac{\bhat}{B},
\label{eq:gamma_gc0}
\end{equation}
so that $\vb{\gamma}_{\rm gc0}[{\cal H}_{\rm gc}]({\bf Z}) = {\bf v}_{\rm gc0}$ and, therefore, energy is conserved. Moreover, since
\begin{eqnarray*}
\vb{\gamma}_{\rm gc0}[{\cal P}_{\rm gc}^{i}]({\bf Z}) &=& \bhat\,b^{i} \;+\; ({\bf B}\btimes\wh{\imath})\btimes\bhat/B \;=\; \wh{\imath}, \\ 
\vb{\gamma}_{\rm gc0}[{\cal P}_{{\rm gc}\varphi}]({\bf Z}) &=& \bhat\,b_{\varphi} \;+\; ({\bf B}\btimes\partial{\bf X}/\partial\varphi)\btimes\bhat/B \;=\; \partial{\bf X}/\partial\varphi, 
\end{eqnarray*}
then both the guiding-center momentum and angular-momentum functionals \eqref{eq:P_gc_i}-\eqref{eq:P_gc_phi} are still collisional invariants. Lastly, we note that the last term in Eq.~\eqref{eq:gamma_gc0} can be removed without jeopardizing the collisional conservation laws because of the locality of the two-particle guiding-center Landau tensor, e.g., $\mathbb{Q}_{\rm gc0}\bdot[\bhat({\bf X})\,b^{i}({\bf X}) - \bhat({\bf X}')\,b^{i}({\bf X}')] \equiv 0$.

\section{Summary}

The present work introduced the metriplectic bracket associated with the guiding-center Vlasov-Maxwell-Landau equations. The guiding-center metriplectic bracket structure is composed of an antisymmetric dissipationless (Hamiltonian) bracket $[\;,\;]_{\rm gc}$, which satisfies the Jacobi property, and a symmetric dissipative bracket $(\;,\;)_{\rm gc}$ that is constructed from the guiding-center Hamiltonian bracket. When the guiding-center entropy functional (which is a Casimir functional of the Hamiltonian bracket) is used in the dissipative bracket, we obtained a Landau representation of the guiding-center Fokker-Planck collisional operator, which guaranteed that the conservations of energy-momentum and angular momentum are automatically satisfied. The guiding-center metriplectic bracket also allowed the self-consistent introduction of collisional effects into the guiding-center Maxwell equation through the guiding-center magnetization and the guiding-center current density.

\vspace*{0.2in}

\acknowledgements

Part of the present work was performed while AJB was visiting the National Institute for Fusion Science (NIFS) in Japan as a Visiting Professor under the Joint Institute for Fusion Theory (JIFT) in the US-Japan Fusion Cooperation Program.  This work was partially supported by the JSPS Grants-in-Aid for Scientific Research (Grant No.\ 24K07000) and by the NIFS Collaborative Research Program (No.\ NIFS23KIPT009). The work by AJB was also supported by the National Science Foundation grant PHY-2206302.

\vspace*{0.2in}

\noindent
{\sf AUTHOR DECLARATIONS}

\noindent
{\bf Conflict of Interest} 

The authors have no conflicts to disclose.

\vspace*{0.2in}

\noindent
{\sf DATA AVAILABILITY}

Data sharing is not applicable to this article as no new data were created or analyzed in this study.

\appendix

\section{\label{sec:Isotropic}Isotropic Field-Particle Model}

In this Appendix, we present a simple model that shows that the guiding-center collisional fluid flow \eqref{eq:coll_gcflow} is diffusive in character\cite{Hazeltine_Meiss}. For this purpose, we introduce the isotropic field-particle Maxwellian model \cite{Brizard_2004}, and we replace the coordinates $(p_{\|},\mu)$ with $(p,\theta \equiv \cos^{-1}p_{\|}/p)$, with Jacobian $p^{2}\sin\theta$, so that the test-particle Fokker-Planck coefficients become
\begin{eqnarray}
{\bf K}_{\rm gc} &=& -\,\nu(p)\,\left(p_{\|}\,\bhat + m\Omega\,\pd{\vb{\rho}_{0}}{\zeta}\right) \;=\; -\nu(p)\,p \left(\cos\theta\,\bhat \;+\frac{}{} \sin\theta\,\wh{\bot}\right) \;\equiv\; -\,\nu(p)\,p\;\wh{\sf p}, \\
\mathbb{D}_{\rm gc} &=& D_{\bot}(p)\,\mathbb{I} \;+\; \left[D_{\|}(p) \;-\; D_{\bot}(p)\right]\;\wh{\sf p}\wh{\sf p},
\end{eqnarray}
where the coefficients $(\nu,D_{\|},D_{\bot})$ are defined in terms of Rosenbluth potentials \cite{Xu_Rosenbluth_1991,Helander_Sigmar,Hirvijoki_Brizard_Pfefferle:2017}, $\mathbb{I}$ denotes the unit dyadic tensor, and $\wh{\bot} \equiv \partial\wh{\rho}/\partial\zeta$ is an explicit function of the gyroangle $\zeta$. Hence, the guiding-center collisional flux \eqref{eq:Sgc_flux} is expressed as
\begin{eqnarray}
{\bf S}_{\rm gc} &=&  -\,\nu\,p\;\wh{\sf p} \;-\; \left[D_{\bot}\,\mathbb{I} \;+\frac{}{} (D_{\|} \;-\; D_{\bot})\;\wh{\sf p}\wh{\sf p}\right] \bdot \left( \wh{\sf p}\,\pd{\ln F}{p} + \frac{\wh{\sf q}}{p}\;\pd{\ln F}{\theta} + \frac{c\bhat}{qB}\btimes\nabla\ln F\right) \nonumber \\
 &=& -\,\nu\,p\;\wh{\sf p} \;-\; D_{\|}\;\pd{\ln F}{p}\;\wh{\sf p} \;-\; \frac{D_{\bot}}{p}\;\pd{\ln F}{\theta}\;\wh{\sf q} \nonumber \\
  &&-\;  \left[D_{\bot}\,\mathbb{I} \;+\frac{}{} (D_{\|} \;-\; D_{\bot})\;\sin^{2}\theta\;\wh{\sf \bot}\wh{\sf \bot}\right] \bdot\frac{c\bhat}{qB}\btimes\nabla\ln F,
\end{eqnarray}
where $\wh{\sf q} \equiv \partial\wh{\sf p}/\partial\theta = -\,\sin\theta\;\bhat + \cos\theta\;\wh{\bot}$ and we omit displaying the guiding-center ordering parameter $\epsilon$. Using the rotating basis $(\bhat,\wh{\bot}, \wh{\rho} \equiv \bhat\btimes\wh{\bot})$, we also find the spatial guiding-center collisional flux
\begin{eqnarray}
\frac{c\bhat}{qB}\btimes{\bf S}_{\rm gc} &=& -\,\frac{c}{qB}\;\left[\left(\nu\,p\;+\; D_{\|}\,\pd{\ln F}{p}\right)\;\sin\theta \;+\; \frac{D_{\bot}}{p}\;\cos\theta\;\pd{\ln F}{\theta}\right]\wh{\rho} \nonumber \\
 &&-\; D_{\bot}\frac{c\bhat}{qB}\btimes\left(\frac{c\bhat}{qB}\btimes\nabla\ln F\right) + (D_{\|} - D_{\bot})\;\frac{c^{2}\sin^{2}\theta}{q^{2}B^{2}}\;\left(\wh{\rho}\frac{}{}\wh{\rho}\right)\bdot\nabla\ln F.
\end{eqnarray}
The guiding-center Fokker-Planck collision operator \eqref{eq:C_gc_FP} 
\begin{equation}
{\cal C}_{\rm gc}[F] \;=\; -\;\frac{1}{p^{2}}\pd{}{p}\left(p^{2}\;\langle S_{\rm gc}^{p}\rangle\;F \right) \;-\; \frac{1}{\sin\theta}\pd{}{\theta}\left(\sin\theta\;\langle S_{\rm gc}^{\theta}\rangle\;F \right) \;-\; \frac{1}{B}\nabla\bdot\left(B\;\langle S_{\rm gc}^{\bf X}\rangle\;F\right) 
\label{eq:gcFP_dk}
\end{equation}
is now expressed in terms of the gyroangle-averaged components
\begin{eqnarray}
\langle S_{\rm gc}^{p}\rangle &\equiv& \left\langle {\bf S}_{\rm gc}\bdot\wh{\sf p}\right\rangle \;=\; -\,\nu\;p \;-\; D_{\|}\;\pd{\ln F}{p}, \label{eq:Sgc_p} \\
\langle S_{\rm gc}^{\theta}\rangle &\equiv& \left\langle {\bf S}_{\rm gc}\bdot\frac{\wh{\sf q}}{p}\right\rangle \;=\; -\;\frac{D_{\bot}}{p^{2}}\;\pd{\ln F}{\theta}, \label{eq:Sgc_theta} \\
\langle S_{\rm gc}^{\bf X}\rangle &\equiv& \frac{c\bhat}{qB}\btimes\langle{\bf S}_{\rm gc}\rangle \;=\; \frac{1}{(m\Omega)^{2}}\left[ D_{\bot} \;+\; \frac{1}{2} (D_{\|} - D_{\bot})\,\sin^{2}\theta \right]\mathbb{I}_{\bot}\bdot\nabla\ln F, \label{eq:Sgc_X}
\end{eqnarray} 
where we used $\langle\wh{\rho}\wh{\rho}\rangle = \frac{1}{2}\,\mathbb{I}_{\bot} \equiv \frac{1}{2}\,(\mathbb{I} - \bhat\bhat)$. In Eq.~\eqref{eq:gcFP_dk}, the first term describes energy dissipation ($\nu)$ and diffusion $(D_{\|})$, the second term
\[ -\; \frac{1}{\sin\theta}\pd{}{\theta}\left(\sin\theta\;\langle S_{\rm gc}^{\theta}\rangle\;F \right) \;=\; \frac{D_{\bot}}{p^{2}}\;\pd{}{\xi}\left[(1 - \xi^{2})\;\pd{F}{\xi}\right] \]
is the standard pitch-angle scattering operator (with $\xi \equiv \cos\theta$), while the third term describes classical spatial diffusion. We note that the drift-kinetic collisional operator \eqref{eq:gcFP_dk} has been shown\cite{Decker_2010} to describe neoclassical transport behavior such as the bootstrap current. In addition, the metriplectic bracket structure introduced in the present work shows how the spatial collisional flux \eqref{eq:Sgc_X} contributes to the collisional guiding-center fluid flow \eqref{eq:coll_gcflow} and the guiding-center magnetization in Eq.~\eqref{eq:Hgc_col}.

\bibliographystyle{unsrt}
\bibliography{gc_Landau}

\end{document}